\documentclass{PoS}

\def\solar{\ifmmode_{\mathord\odot}\else$_{\mathord\odot}$\fi~}
\def\gsim{\stackrel{>}{_\sim}}
\def\lsim{\stackrel{<}{_\sim}}
\newcommand{\n}{\noindent}

\title{VLBI at the highest frequencies -- AGN studied with micro-arcsecond resolution}

\ShortTitle{VLBI at the highest frequencies}

\author{\speaker{Thomas P. Krichbaum}\\
        Max-Planck-Institute f\"ur Radioastronomie, Bonn, Germany\\
        E-mail: \email{tkrichbaum@mpifr-bonn.mpg.de}}

\author{I. Agudo\\
        Max-Planck-Institute f\"ur Radioastronomie, Bonn, Germany\\
        E-mail: \email{iagudo@mpifr-bonn.mpg.de}}

\author{U. Bach\\
        Max-Planck-Institute f\"ur Radioastronomie, Bonn, Germany\\
        E-mail: \email{ubach@mpifr-bonn.mpg.de}}

\author{A. Witzel\\
       Max-Planck-Institute f\"ur Radioastronomie, Bonn, Germany\\
       E-mail: \email{awitzel@mpifr-bonn.mpg.de}}

\author{J. A. Zensus\\
       Max-Planck-Institute f\"ur Radioastronomie, Bonn, Germany\\
       E-mail: \email{zensus@mpifr-bonn.mpg.de}}

\abstract{Compact galactic and extragalactic radio sources can be imaged with 
an unsurpassed angular resolution of a few ten micro-arcseconds, adopting the 
observing technique of global millimeter VLBI. Here we present the Global Millimeter VLBI 
Array (GMVA) and discuss its present performance. For individual and partially archetypical
radio sources with prominent VLBI jets (e.g. 3C\,120, Cygnus\,A, M\,87, 3C\,454.3, NRAO150), 
we show and discuss new results obtained with the GMVA. The variety of observed effects 
range from jet propagation and bending, 
partial fore-ground absorption in the nucleus, and jet component ejection after major flares 
to new and very small ($15-20$ Schwarzschild radii) upper limits to the jet base of M\,87.
We also discuss the future development of mm-VLBI at 3\,mm and towards shorter wavelengths 
and make suggestions for possible improvements.}

\FullConference{The 8th European VLBI Network Symposium on New Developments in VLBI Science and Technology
                and EVN Users Meeting  \\
		September 26-29 2006\\
		Torun, Poland}

\begin{document}

\section{Introduction}

VLBI studies at centimeter wavelengths have already revealed a wealth of details about the morphology 
and propagation of the powerful AGN radio jets. The detailed understanding of their physical origin, however,
still is limited and it is very unclear, how these jets are launched and accelerated and how the jet base 
is connected to the `central engine'. The recent development in black hole theory and computer 
simulations suggest that general relativistic magneto-hydrodynamical (GR-MHD) processes, i.e. the interplay 
of matter accretion, magnetic fields and space-time curvature, eventually modified by black hole
rotation (Kerr metric), may play an important role (i.e. \cite{Gammie03}, \cite{deVilliers03}, \cite{Hawley06}). 
However, not many observational constraints to the physical 
parameters of these models are yet available. It is also unclear, if GR-MHD driven jet acceleration occurs 
in all classes of radio-loud AGN, or is present in only some objects.
It is hoped that VLBI studies of AGN at short millimeter wavelengths
($\lambda \leq 3$\,mm, mm-VLBI) can contribute in answering these questions. Depending on the physical details of the 
jet acceleration and collimation processes, the compactness and brightness temperatures of the radio cores 
must change with distance from the center. Similarly, the morphological, spectral and kinematical properties 
of the inner-most jet regions will change, depending on the nature of the `central accelerator'.  
VLBI imaging at the highest possible frequencies therefore should be used to determine/measure the variation of 
observational parameters, like i.e. the brightness temperature, jet-velocity and jet-spectrum along the jet and
as close as possible down to the jet origin.
Since at cm-wavelengths, the jet base appears self-absorbed and unresolved, 
the highest possible observing frequencies,
and angular resolutions are required to image and study these regions. 

\begin{figure}[t]
\hspace*{5mm} \includegraphics[width=0.45\textwidth,angle=-90]{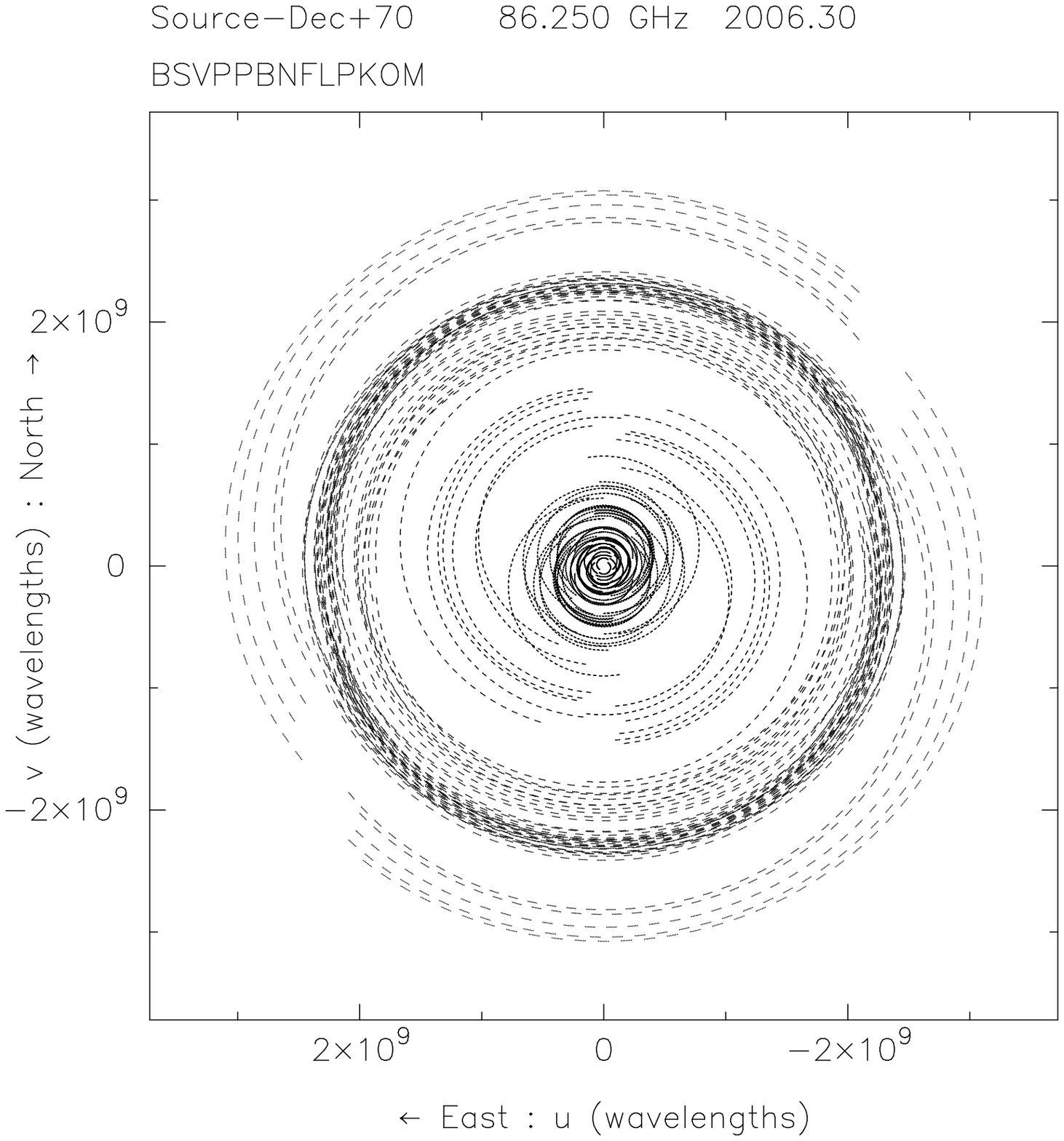}
\hspace*{5mm} \includegraphics[width=0.45\textwidth,angle=-90]{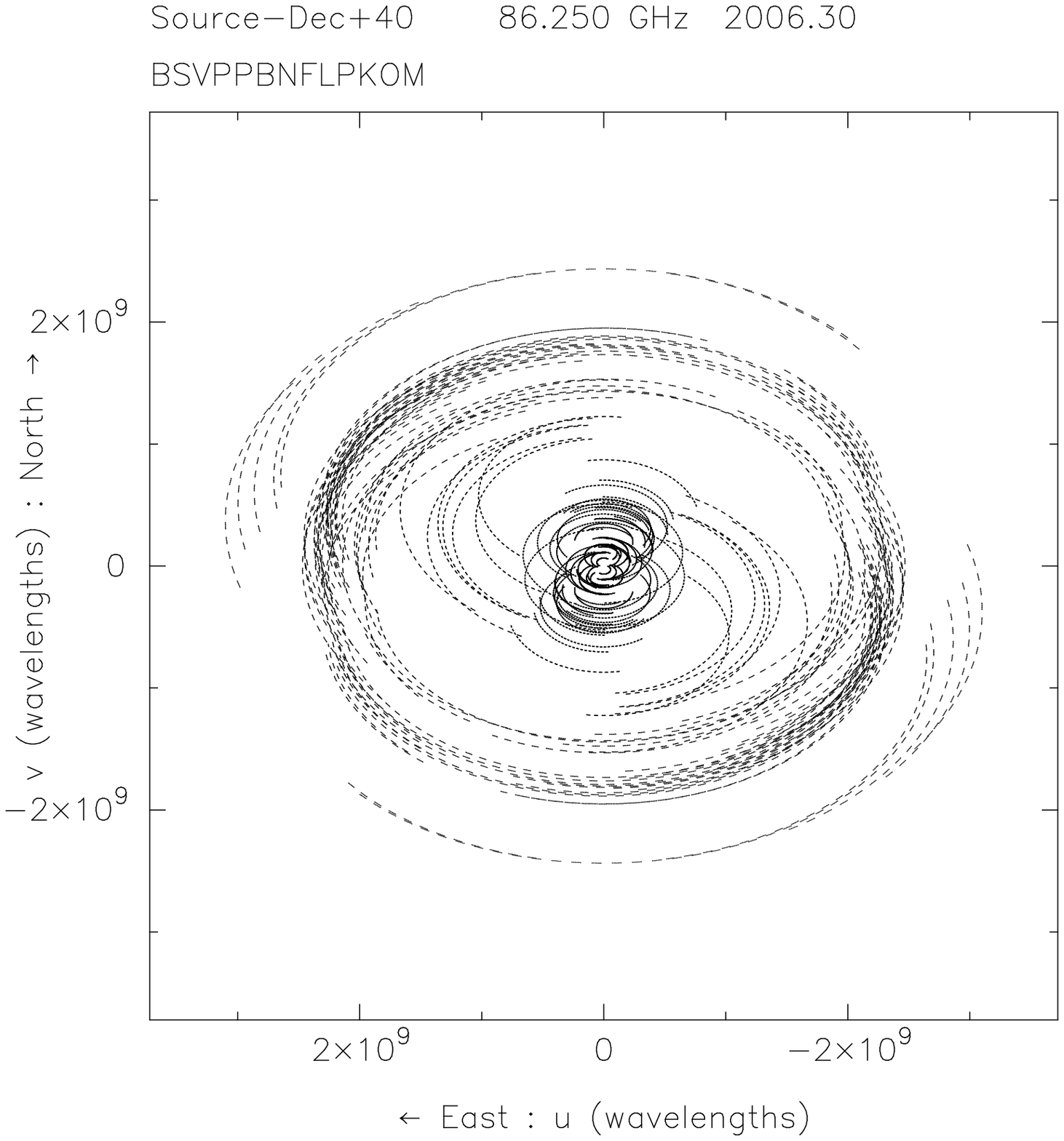} \\
~~~\\
\hspace*{5mm} \includegraphics[width=0.45\textwidth,angle=-90]{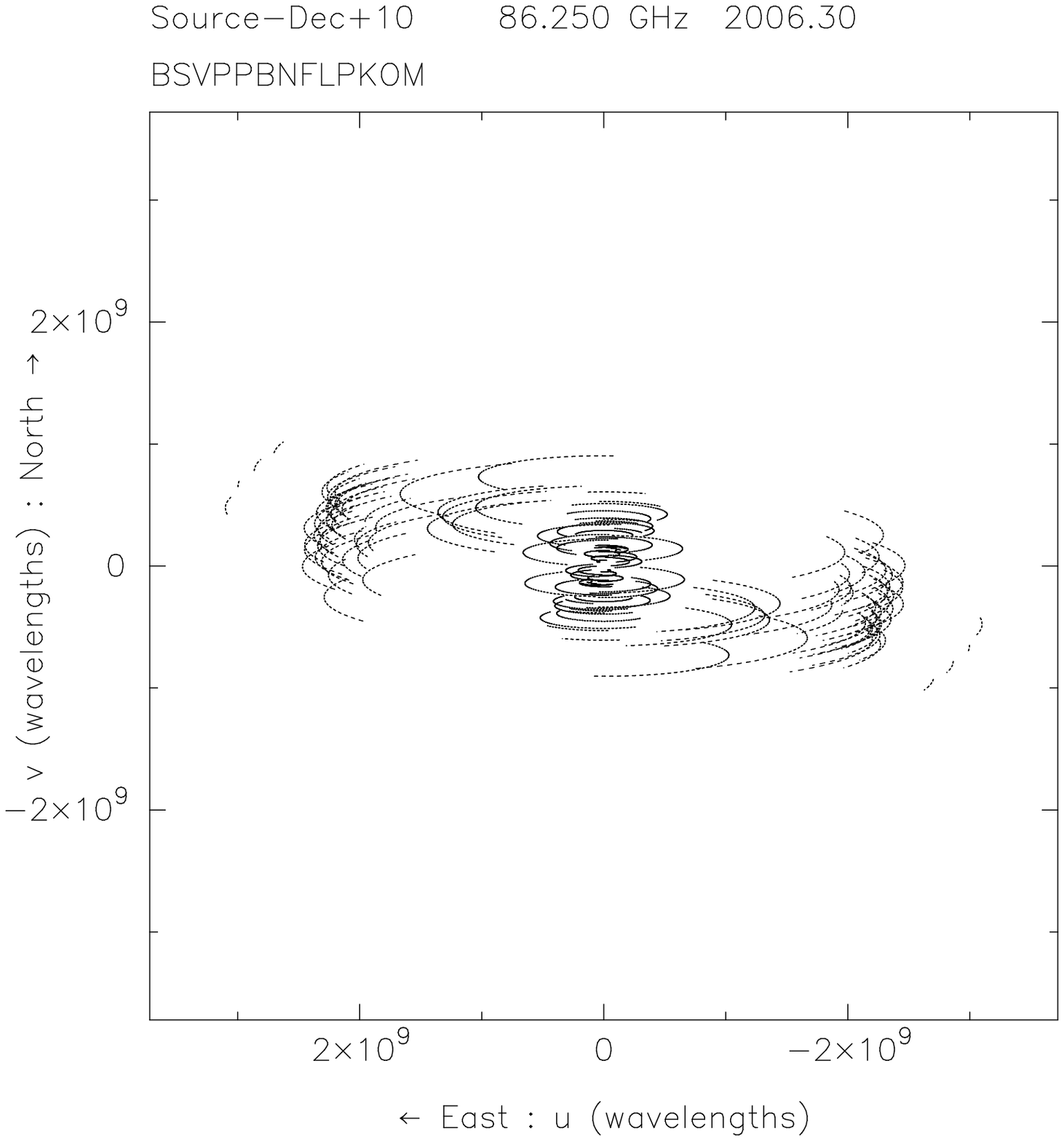}
\hspace*{5mm} \includegraphics[width=0.45\textwidth,angle=-90]{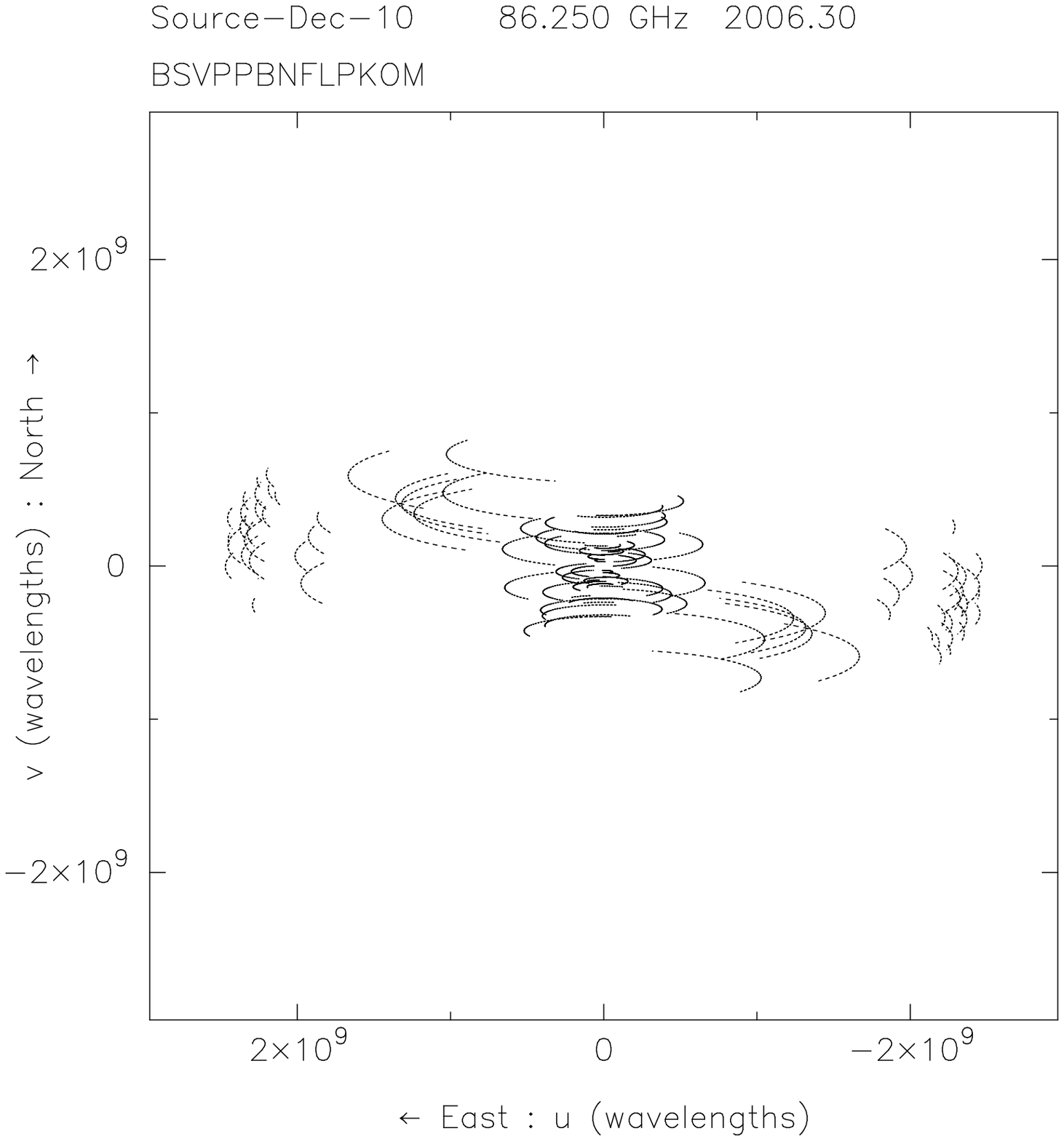}

\caption{Typical uv-coverages for the VLBI array of the 13 radio telescopes presently participating in the global 3mm VLBI
with the GMVA. For each observation a duration of 12 hrs and a duty cycle (VLBI time over total time)
of $\sim 50$\,\% was adopted.
From top left to bottom right, uv-coverages for a source at declination $\delta=70^\circ$, $40^\circ$,
$10^\circ$, and $-10^\circ$ are shown.
}
\label{uvplot}
\end{figure}

\section{Present day Global Millimeter VLBI}

We now will concentrate on some new results obtained with the 
Global mm-VLBI Array (GMVA)\footnote{web link: http://www.mpifr-bonn.mpg.de/div/vlbi/globalmm}.
The GMVA is open to the scientific community, offering regular global VLBI observing in the 3\,mm band.
At present, two observing sessions per year, each of up to 5 days duration are performed.
More frequent observations are in principle possible, if the demand for 
this (via proposal pressure) increases.
Proposal deadlines are synchronized with the annual February 1st and October 1st deadlines of the NRAO.
The GMVA combines the European antennas (the 100\,m Effelsberg (EB), Germany, the 30\,m Pico Veleta (PV), Spain, 
the phased 6x15\,m Plateau de Bure Interferometer (PdB), France, the 20\,m Onsala (ON), Sweden , and the 
14\,m Mets\"ahovi (MET), Finland) with the VLBA, of which eight antennas are presently 
equipped with 3\,mm receivers. In Figure \ref{uvplot}, we show some typical uv-coverages
for the GMVA for sources at 4 different representative declinations. Mainly due to the participation of the two
very sensitive IRAM telescopes (PV \& PdB) and the 100\,m Effelsberg telescope, the array sensitivity is 
improved by a factor of $3-4$, when compared to the VLBA alone. In the standard observing mode
with Mark 5 recording at 512\,Mbit/s, the single baseline $7 \sigma$--detection threshold
on the most sensitive PV--PdB baseline is 30-40\,mJy, and from the IRAM antennas to one VLBA antenna
it is 50-100\,mJy (assuming 100\,sec integration and 20\,sec coherence time). This has to be compared
with a $200-300$\,mJy detection level, which is reached between two VLBA antennas. The inner-American
sensitivity will improve, when the 100\,m Green Bank telescopes (GBT) becomes available for 3\,mm VLBI.
The quality of present day 3\,mm VLBI maps and their rms noise level, is mainly determined by the 
atmospheric conditions during the observations. In a compromise for good observing conditions at all
VLBI stations, annual observations in spring (April, May) and autumn (October) are performed. 
It is frequently discussed, if mm-VLBI observations during winter time (January, February) might be better suited. However,
cold periods of low humidity appear often only locally and are neither well predictable nor are in present European climate
long enough. With up to 13 stations participating in a GMVA observation, 
images with a dynamic range of up to a few hundred are obtained, reaching a rms noise level of $\sim 0.5-2$\,mJy/beam 
for a 12 hr full uv-coverage observation (at 512\,Mbit/s, 50\,\% duty cycle).
We note that a relatively large number of participating VLBI stations plays an important role for the imaging, 
because the short coherence time and the atmospheric opacity variations require the heavy use of closure 
amplitudes and require amplitude self-calibration down to the time scale of atmospheric fluctuations. 
In radio interferometry it is well known that self-calibration only
leads to reliable images, if the number of participating telescopes is large enough ($N \geq 10$). 

For compact galactic and extragalactic radio sources, the GMVA provides VLBI images with an angular 
resolution of up to $40$\,$\mu$as. For a source at redshift $z=0.1$, this corresponds to a linear scale of
$2.2 \cdot 10^{17}$\,cm (or 0.07\,pc or 85 light days)\footnote{Through out this paper we adopt the following 
cosmological constants: $H_0 = 71$\,km/s/Mpc, $\Omega_M=0.27$, $\Lambda=0.73$.}. When expressed in terms of Schwarzschild
radii of a black hole of mass $10^9$ M\solar ($R_S^9:= R_S(10^9 {\rm M\solar})$), a linear size of 740\,$R_S^9$ is obtained, comparable
to the expected size of an accretion disk. Mainly technically driven VLBI pilot experiments, performed at 2\,mm and 1\,mm
revealed recently first transatlantic fringes at a world record fringe spacing 
of up to $\sim 6$ G$\lambda$ ($30-35 \mu$as) (for details see \cite{Kri02}, \cite{Kri05}, \cite{Doele05}).
This demonstrates the principal feasibility of global VLBI at wavelengths shorter than $\lambda \leq 2$\,mm, 
and shows that VLBI imaging with $15-20$\,$\mu$as resolution is not an unrealistic dream.  Depending on source distance and
mass of the central black hole, this corresponds to a few to a few hundred gravitational radii (or Schwarzschild radii) 
in size. This resolution is so high that future and more sensitive mm-VLBI experiments at $\lambda \leq 1$\,mm, should 
facilitate the observation of direct signatures of General Relativity effects near super massive black holes (SMBH).

\begin{figure}
\begin{minipage}[t!]{0.49\textwidth}{ \hspace*{10mm}
\includegraphics[bb=35 135 530 675,clip,width=0.8\textwidth,origin=c,angle=-90]{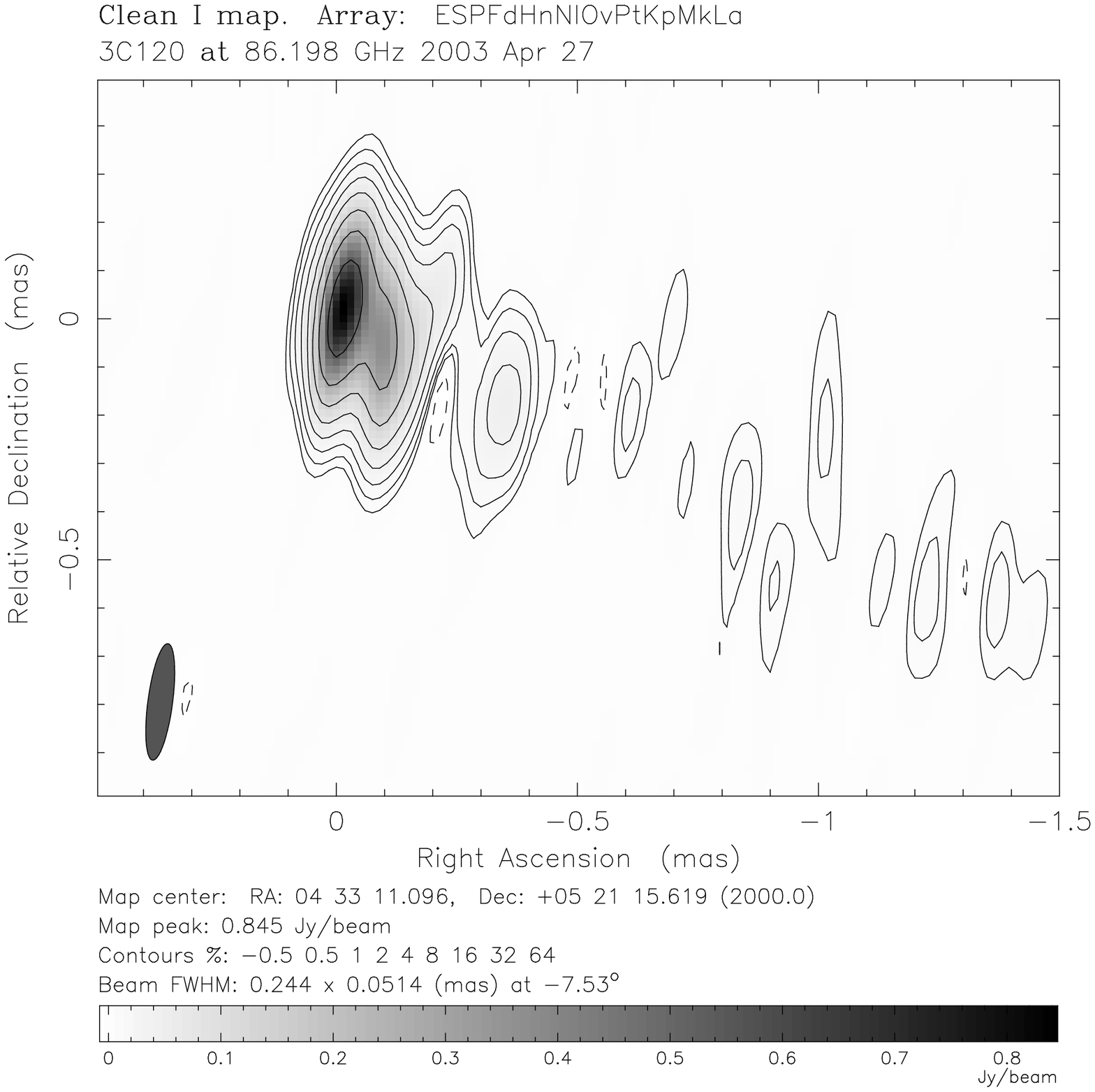}

\vspace*{3mm}
\hspace*{10mm}
\includegraphics[bb=35 145 530 666,clip,width=0.83\textwidth,origin=c,angle=-90]{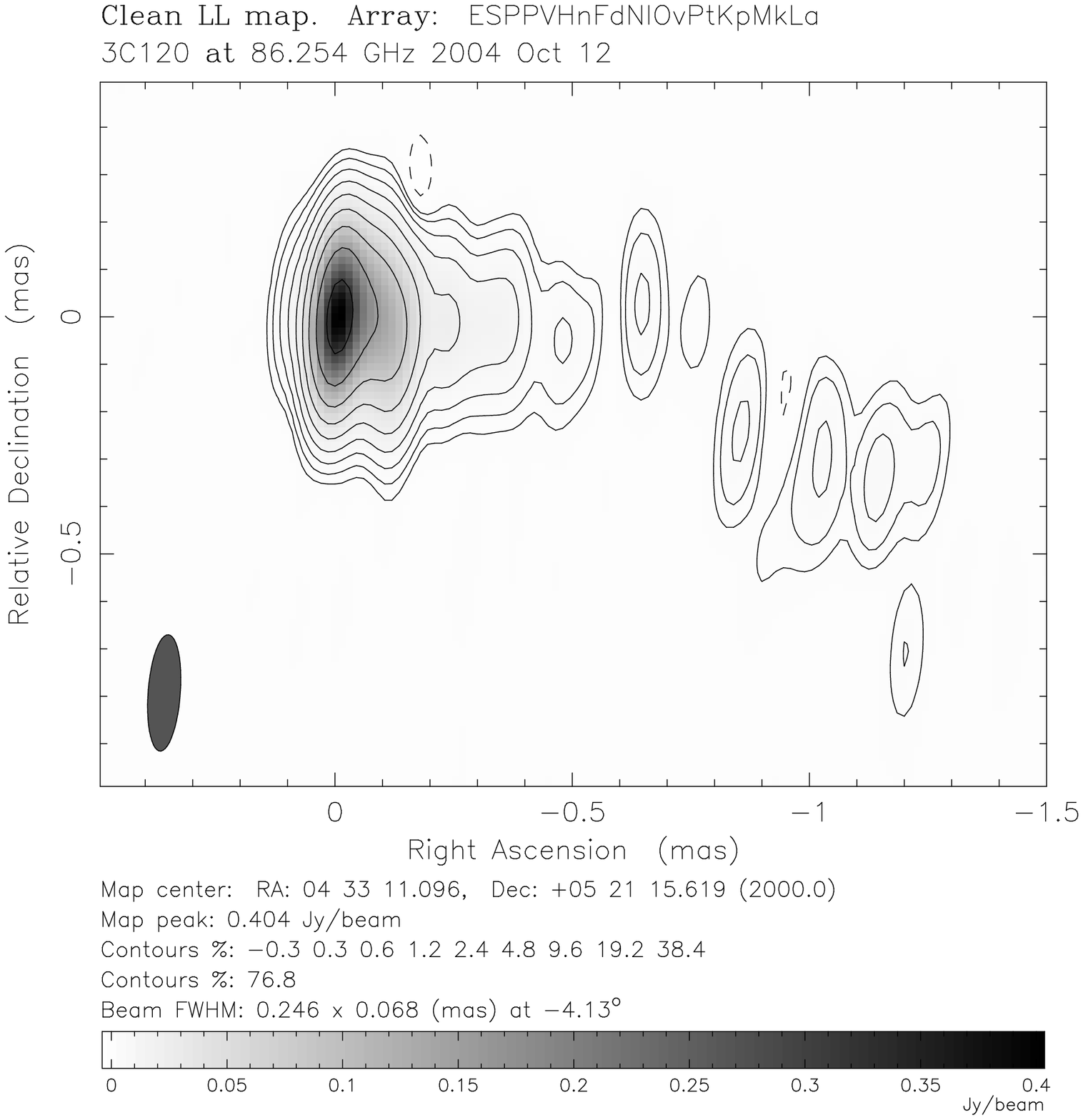}
}
\vspace*{-15mm}
\end{minipage}
~~~
\begin{minipage}[t!]{0.49\textwidth}{ \hspace*{10mm}
\includegraphics[width=0.8\textwidth,angle=0]{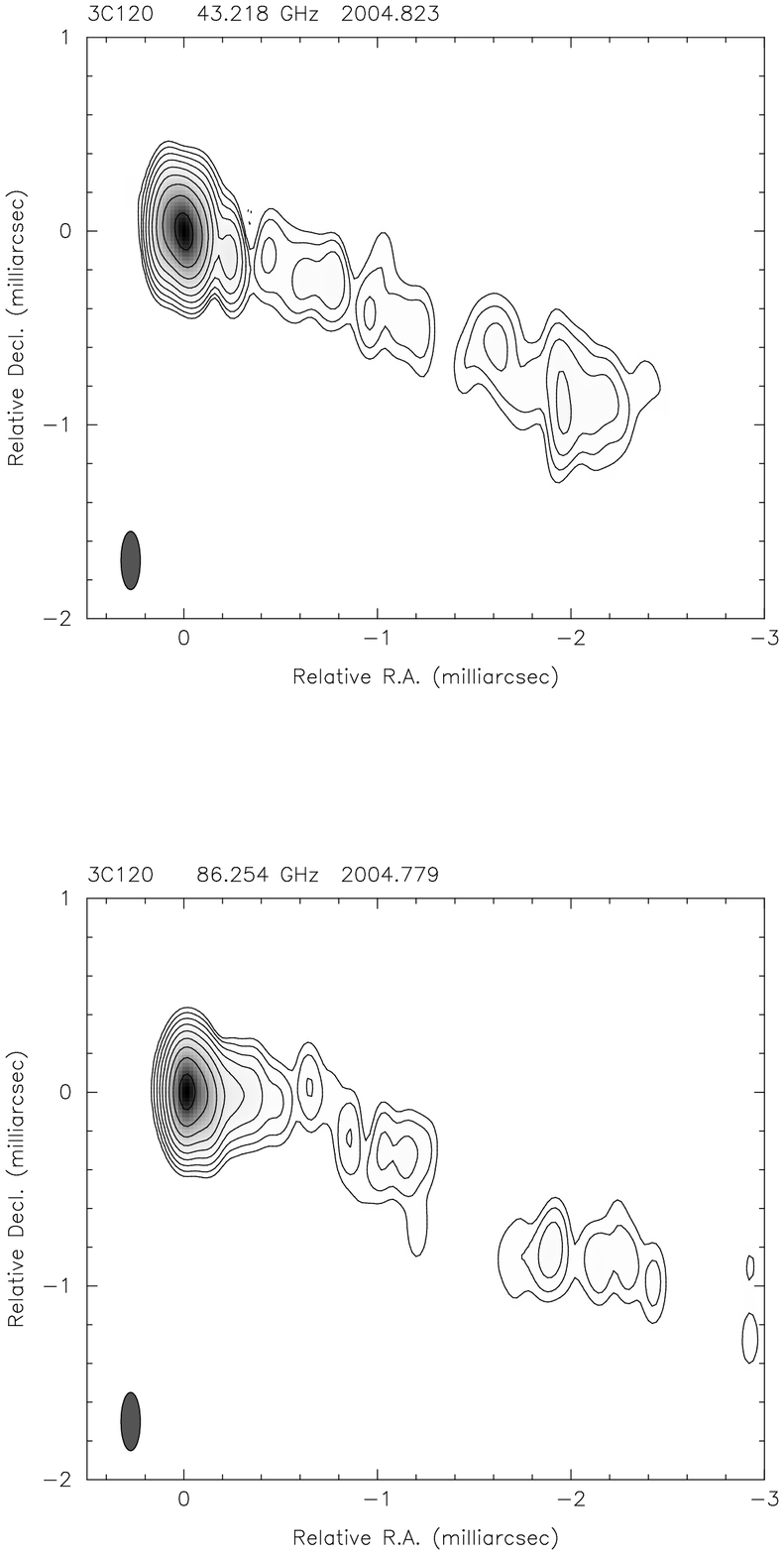}
}
\end{minipage}

\vspace*{10mm}
\caption{{\bf Left:} Clean maps of 3C\,120 observed at 86\,GHz with the GMVA in April 2003 (top),
and October 2004 (below). Contours and peak fluxes are shown below each map.
{\bf Right:}
3C\,120 observed near in time at 43\,GHz with the VLBA (top) and at 86\,GHz with the GMVA (below).
The 86\,GHz image is from October 12, the  43\,GHz image is from October 28, 2004 (Marscher
et al., \cite{Marscher07}). Contour levels are at -0.3, 0.3, 0.6, 1.2, 2.4, 4.8, 9.6, 19.2, 38.4,
and 76.8\,Jy/beam with peak flux of 0.73\,Jy/beam at 43\,GHz, and 0.46\,Jy/beam at 86\,GHz.
For better comparison, both maps are convolved with the same elliptical beam of size $0.3 \times 0.1$\,mas.
}
\label{3c120}
\end{figure}

\subsection{Examples of Radio-Jets studied with 3\,mm-VLBI:}

\n
In the following section we highlight some new and partly still preliminary results obtained from 
recent global 3\,mm VLBI observations. The results of a more statistically oriented new 3\,mm VLBI detection
survey are presented by Lee et al. (this conference). \\

\n
{\bf 3C\,120:}
In Figure \ref{3c120} we show two VLBI maps of 3C\,120 obtained with the GMVA in April 2003 (10 stations, mode 256-8-2, 11 hrs)
and in October 2004 (12 stations, mode 512-8-2, 12 hrs). The dynamic ranges of the two maps are 200:1 and 300:1, respectively.
At the redshift of $z=0.033$ the observing beam size of $\sim 0.25 \times 0.06$\,mas corresponds to a spatial
resolution of $0.16 \times 0.04$\,pc, or in terms of a $3 \cdot 10^7$\,$M\solar$ black hole
to a linear scale of $4.1 \times 10^4$ $R_S^7$ for the minor beam axis. At and below the 0.5\,mas scale,
structural differences and a misalignment of the orientation of the inner jet between both epochs are visible. 
Much denser time sampled observations will be necessary to follow the expected motion of $\sim
2$\,mas/yr (\cite{Gomez01}). Slightly super-resolving the map of April 2003, reveals an 'S-shaped' structure on the
$0.1-0.2$\,mas scale. This and the change of the position angle of the sub-mas jet on a 1.5\,yr time scale 
supports the idea of helical Kelvin-Helmholtz instabilities, which may be excited by some sort 
of `rotation' at the jet base (\cite{Caproni04}, \cite{Hardee05}). 

Beyond the 0.5\,mas scale, the 3\,mm VLBI jet fades and breaks into filamentary sub-components. A demonstration
of the reality of this weak and extended jet emission is obtained from the inspection of the right panel of 
Figure \ref{3c120}. Here, we show uv-tapered VLBI images observed within a 16\,day interval at 43\,GHz (VLBA) and 86\,GHz (GMVA). 
Both maps are convolved with the same CLEAN-beam of $0.3 \times 0.1$\,mas in size, and show weak 
jet emission at corresponding locations on the mas-scale. A future and more detailed analysis will be necessary to find out, 
if the partial misalignment of the jet ridge-line in $0.2-0.5$\,mas region can be interpreted by transverse 
opacity effects and jet stratification. \\

\begin{figure}
\includegraphics[angle=-90,width=0.98\textwidth]{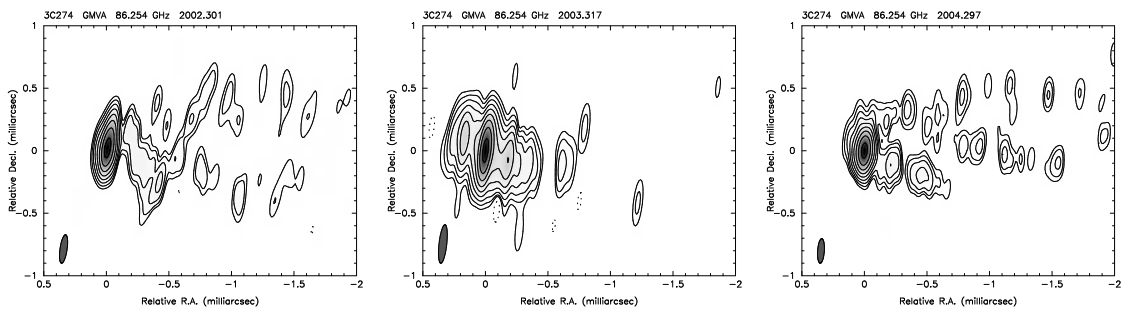}
\caption{Preliminary 86\,GHz VLBI images of M\,87 (3C\,274) obtained with global 3\,mm VLBI
in April 2002, 2003, and 2004 (from left to right).
The contour levels are at -0.3, 0.3, 0.6, 1.2, 2.4, 4.8, 9.6, 19.2, 38.4, and 76.8 \% of the peak flux, of
0.52 Jy/beam (2002), 0.73 Jy/beam (2003), 0.35 Jy/beam (2004). In the map of 2002, the lowest contour
level is omitted. The restoring beam sizes are $0.23 \times 0.057$\,mas (2002), $0.31 \times 0.062$\,mas (2003),
and $0.20 \times 0.054$\,mas (2004), respectively.
\newline
}
\label{m87}
\end{figure}

\n
{\bf M\,87:} The nearby radio galaxy M\,87 (Virgo\,A, 3C\,274) is one of 
the closest radio galaxies with a prominent radio jet (z=0.00436, D=16.75\,Mpc).
It is the ideal object for high spatial resolution VLBI studies, which address
the question how the powerful and relativistic jets in AGN are made. 
For the assumed $3 \cdot 10^9 M\solar$ black hole, the Schwarzschild radius is $R_S^9 = 
8.9 \cdot 10^{14}$\,cm, which translates for a GMVA observing beam of $50-60$\,$\mu$as
into an exceptionally high spatial resolution corresponding to only $14-17$ Schwarzschild radii for this source.

We note that a similar spatial resolution in terms of Schwarzschild radii is obtained for
the $3.6 \cdot 10^6$\,$M\solar$ black hole in the center of our Galaxy (Sgr\,A*) at the shorter wavelength of 1.3\,mm
(for details see \cite{Kri98}, \cite{Kri07} and references therein).
The low declination of Sgr\,A* and the lack of sensitive millimeter antennas in the Southern
Hemisphere, however, will most likely not permit to image Sgr\,A* at 1\,mm with sufficiently high dynamic range
(and a not too elongated observing beam) for the next $5-10$ years. In this context, the mm-VLBI observations 
of M\,87 are more promising for the search of GR-effects in the vicinity of a SMBH. 

Since the mid 1990's, several attempts were made to obtain a reliable image of M\,87 with global 3mm VLBI.
With a correlated flux density of only $\sim 300-600$\,mJy at the longest uv-spacings ($\sim 3$\,G$\lambda$),
the nucleus of M\,87 is relatively weak and fringe detection requires a high sensitivity, which is now provided
by the GMVA. 
The preliminary maps presented in Figure \ref{m87} show the structure of the inner jet at three epochs: April 2002,
2003 and 2004, respectively. The observations in 2002 and 2003 were made at 256\,Mbit/s, while in 2004 the recording
rate was increased to 512\,Mbit/s. The amount of visibility data with the three most sensitive telescopes (EB, PV, PdB)
largely determines the quality of the maps.

In April 2002, the source was observed with 12 VLBI stations (8 x VLBA, EB, ON,
PV, HSTK). Unfortunately bad weather at EB and PV limits the dynamic range of the resulting map (Fig. \ref{m87}, left).
Tapering this image with a larger beam (not shown here), however, shows an edge brightened transversely resolved
and conical jet with similar opening angle as seen in the 43\,GHz images of the
VLBA (\cite{Junor99}, \cite{Ly2004}) and also in the later 3\,mm map of April 2004 (Figure \ref{m87}, right). 

In April 2003, the source was observed with participation
of the phased IRAM interferometer on Plateau de Bure. The combination of this sensitive telescope
with the other VLBI stations (EB, PV, ON, 8 x VLBA), resulted in a better map (see Fig. \ref{m87}, center).
The source exhibits a core-jet structure and embedded, several distinct emission features, which are aligned 
at a position angle of $pa \sim -(110 - 120)^\circ$, which coincides with the orientation of the southern
border of the jet cone (cf. Fig. \ref{m87}, right). The two times higher brightness of the VLBI core in this epoch (peak flux densities
of 0.52\,Jy/beam in 2002, 0.73\,Jy/beam in 2003, and 0.35\,Jy/beam in 2004) and the dynamic range limitations
may explain the lack of faint jet emission, seen at larger core separations in the 2002 and 2004 image.
In 2002 and 2004, the jet appears one-sided with all of the extended jet emission located west of the brightest
component. In 2003, however, a partially resolved component located $\sim 0.2$\,mas east to the $\sim 4$ times 
brighter central and more compact component is seen.  At present it is unclear, if this eastern
component should be identified with the jet base (which then must vary considerably in compactness),
or if it is part of a counter-jet. Independent evidence for a counter-jet comes from 
a new 43\,GHz VLBA map obtained by Ly et al., \cite{Ly2004}, but is not seen at cm-wavelengths (\cite{Dodson06}).

In April 2004, M\,87 was observed with 10 stations (EB, ON, PV, 7x VLBA). Due to the failure of its H-maser, 
the PdB interferometer was not available (a new H-maser is now delivered). 
Because of the now increased recording rate of 512\,Mbit/s, the
imaging sensitivity was improved and the source could be reliably imaged. The map of Figure \ref{m87} (right) shows 
again the known Y-shaped (conical) jet expansion, most of which happens on the $\leq 0.3 - 0.5$\,mas scale.  
Thus, the linear size of this region of jet collimation is $\sim (85 - 140)$ Schwarzschild radii.
The transverse width of the jet at $r=0.5-2$\,mas is of order of $0.5-0.7$\,mas, corresponding to $\sim 140-200$
Schwarzschild radii, showing clear signs of edge-brightening and a `hollow' or at least faint central jet body.
If this central jet region would contain a fast electron population (fast spine), the $\geq 20-40^\circ$ jet inclination
relative to the observer could lead to Doppler de-boosting, which would reduce its observable brightness.
In 2004 (and in 2002) the jet base appears unresolved, with an upper limit to its size determined by the actual
observing beam. If we assume that the minor axis of the observing beam is a measure of the underlying 
size of the foot-point of the jet, we obtain an upper limit of $54 \mu$as, which corresponds to a spatial scale of only 
$\leq 15$\,$R_S^9$. A lower limit of the brightness temperature of the core component then 
is $T_B \geq 2 \cdot 10^{10}$\,K.
No strong statement can yet be made with regard to the distance of the jet base 
relative to the central super massive black hole. The back-extrapolation of the jet cone to its vertex, however, indicates that 
this distance is small and of order of $\leq 0.1-0.3$\,mas, or $\leq 30-90$ $R_S^9$. 

We find it very remarkable that the size of the VLBI core, which should be identified with the region of the
jet, where it first becomes radiative, is so small. The existence of a fully developed jet
on such small spatial scales will give important new constraints for the theory of jet formation.
Jet models which are based on the magnetic sling shot mechanism (Blandford \& Payne, \cite{BlandfordPayne82}, 
\cite{Cam90}) assume efficient 
particle acceleration along the S-shaped magnetic field lines up to Lorentz factors of $\gamma= 10-20$. 
The field lines are anchored in the inner part of the rotating accretion disk and
probably form helical magnetic flux tubes, which may explain the observed bent and sometimes edge-brightened jet 
morphology (core-sheath structure). In these models, the diameter of the jet base
is defined by the transverse width of the jet after initial acceleration, which is
of order $\geq \gamma \cdot 2R_L$, where $\gamma$ is the bulk Lorentz factor and $R_L$ 
is the radius of the light cylinder (\cite{Cam90}, \cite{TomimatsuTakahashi03}). 
The light cylinder should have a radius of typically $(10-50)\,R_S$ (\cite{Cam90},\cite{FendtMemola01}). 
If we assume for the Lorentz factor $2 \leq \gamma \leq 3$  (as derived from the motion seen
on pc scales), one would expect a size of the jet base of $\geq  (40 - 300) R_S^9$,
which is at least 3 times larger than the measured size. One way out of the problem is to assume 
that the central BH is rotating (Kerr BH). This would reduce the radius of the light cylinder 
and the width of the jet base by up to a factor of 3 (e.g. \cite{Meier01}). In this case, 
the observed small size of the VLBI core would point more towards jet models, in which the jet 
gains energy directly from BH-rotation (\cite{BlandfordZnajek77}, \cite{McKinney06}, \cite{Hawley06}). \\

\begin{figure}
\begin{minipage}[t!]{0.49\textwidth}{
\centering \includegraphics[width=0.65\textwidth,angle=-90]{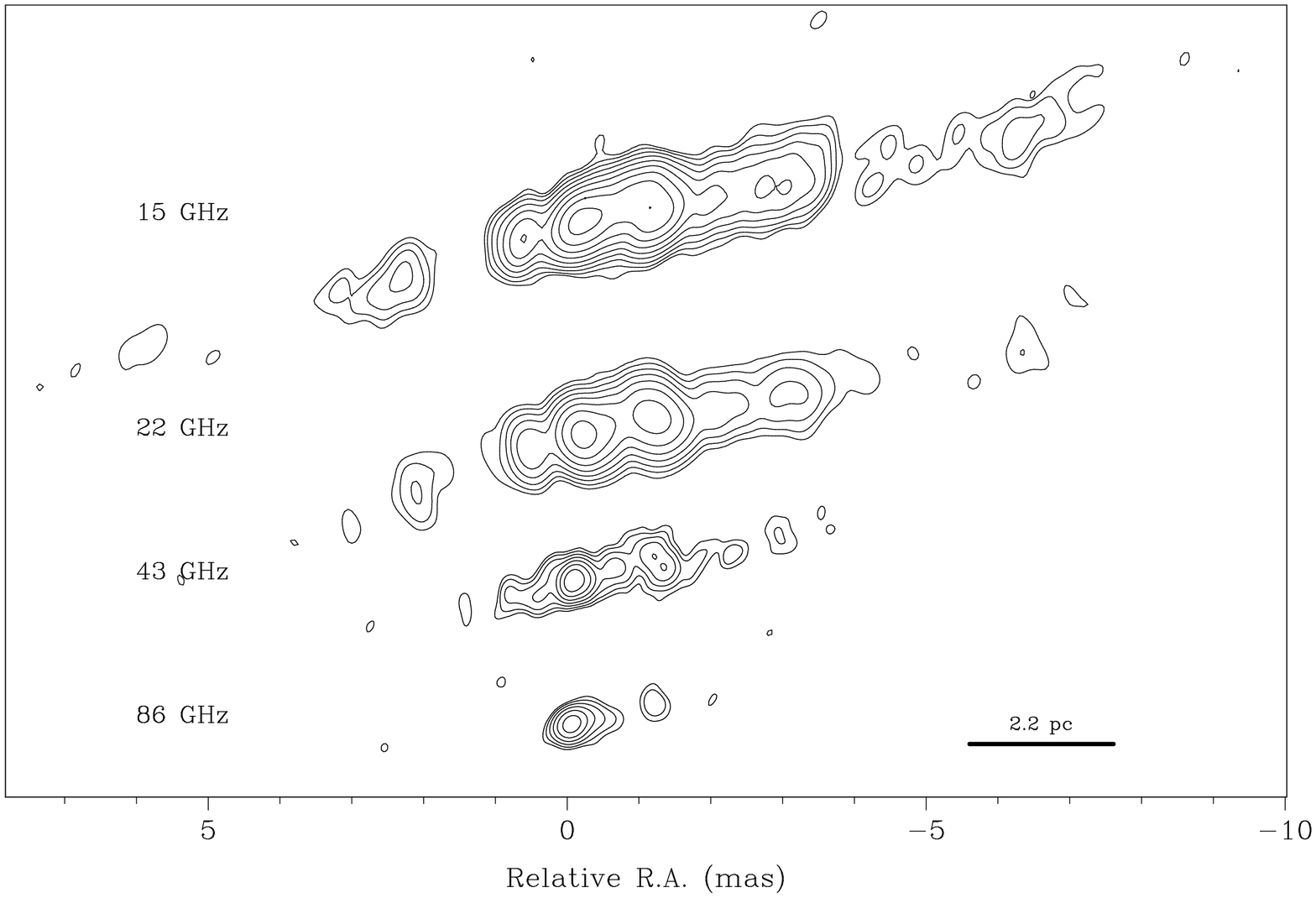}
}
\end{minipage}
~~~~
\begin{minipage}[t!]{0.49\textwidth}{
\includegraphics[width=0.9\textwidth,angle=0]{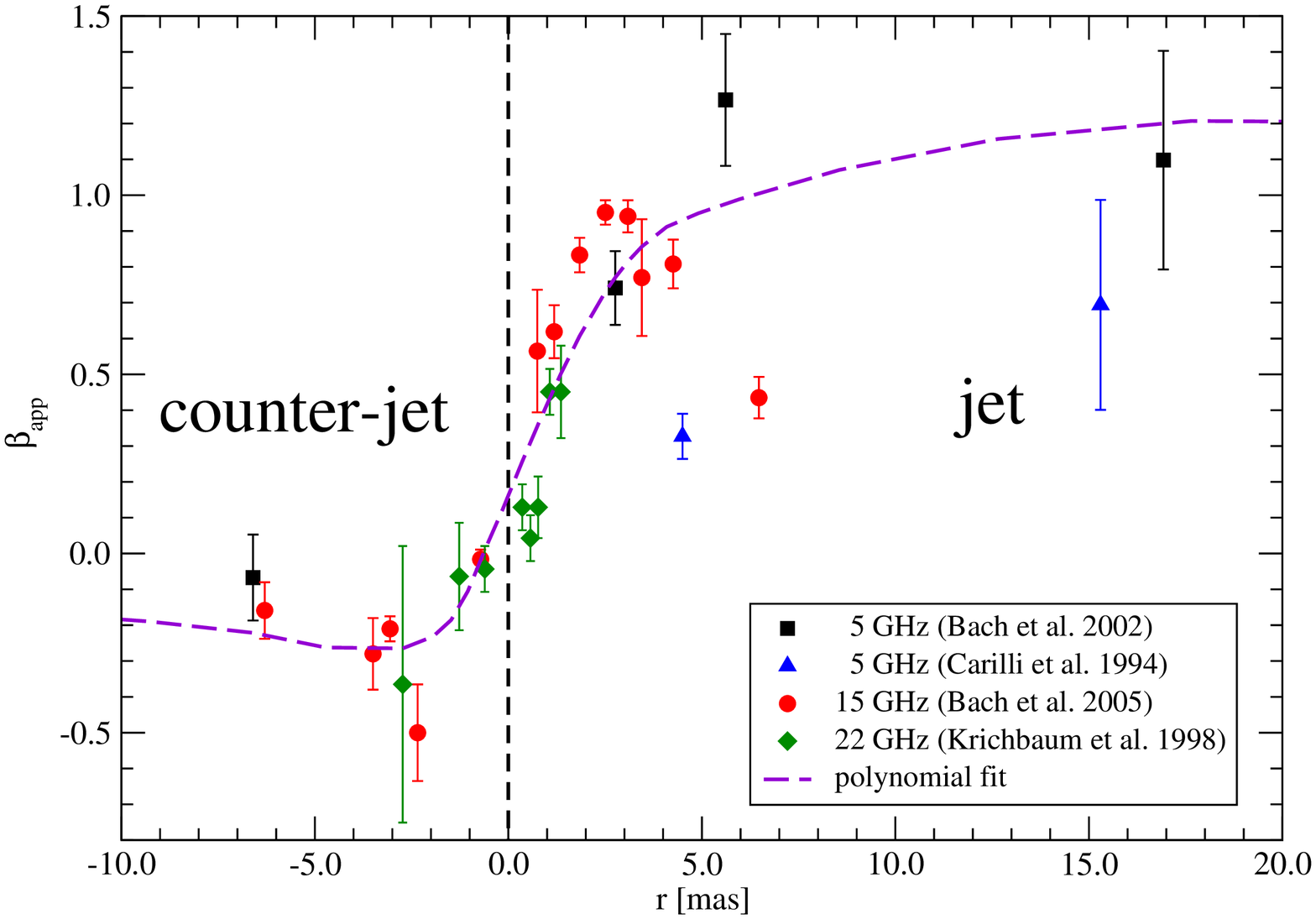}
}
\end{minipage}
\caption{{\bf Left:} The jet of Cygnus\,A at 4 different frequencies: 15, 22, 43, and 86\,GHz (top to bottom).
The 86\,GHz images results from GMVA observations of April 9, 2003 (using 256\,Mbps recording).
The 15 and 22\,GHz observations are phase-referenced, the alignment of the other maps is done based
on morphological similarities.
At 86\,GHz, the beam size is $0.07 \times 0.28$\,mas$^2$ at $-23\,^\circ$, the peak flux density
is 0.33\,Jy and the rms noise is 5.4\,mJy. The lowest contours start at 15\,mJy/beam and
increases in steps of 2. {\bf Right:}
Apparent velocity of individual jet components in jet and counter-jet of Cygnus\,A,
plotted versus core separation. New data are combined with data from the literature.
The dashed line is a polynomial fit including also data points at larger core-separations,
which are not shown in the figure. For details, see Bach et al., \cite{Bach04} \& \cite{Bach05}, 
and references therein.
}
\label{cyga}
\end{figure}

\n
{\bf Cygnus\,A:} Another argument for the relevance of magnetic acceleration in radio jets comes
from multi-frequency studies of the kinematics in the jet and counter-jet of Cygnus\,A. In Figure
\ref{cyga} (left) we show some VLBI maps of the source at $15-86$\,GHz (data: Bach et al., see \cite{Bach04}, \cite{Bach05}). 
At 86\,GHz a one sided core-jet structure is visible, consisting of an unresolved core of $\sim 70$\,$\mu$as size 
(linear size $\leq 90$\,light-days) and two secondary jet components. Some station failure and a recording rate
of only 256\,Mbit/s limit the sensitivity and dynamic range of the map, and inhibit the detection of fainter 
jet and counter-jet emission.  We note that the strong frequency dependence of the jet-to-counter-jet 
ratio suggests the existence of a fore-ground absorber, which attenuates the radiation from the 
counter-jet (\cite{Kri98cyg}). Clearly more sensitive mm-VLBI observations are required, to detect
the counter-jet at a frequency, where the absorber becomes optically thin and the counter-jet becomes brighter. Future high  
angular resolution images of double-sided jets, like in Cygnus\,A, are very important, as they may lead to
the detection of a gap between the two foot-points of jet and counter-jet. The size and position of
the gap would nicely constrain the position of the central BH, and tell us more about the 
jet acceleration processes, which link the BH with the jet base. That magnetic acceleration is not just
speculation is demonstrated in Figure \ref{cyga} (right), which shows the measured apparent jet speed as
a function of core separation. For an almost perfectly straight jet like in Cygnus\,A, jet bending and geometrically
caused acceleration (due to differential Doppler boosting) 
can be neglected. Thus, the observed acceleration must be related to an intrinsic variation 
of the bulk Lorentz factor. The observed large and systematic variation in apparent jet speed 
by almost one order of magnitude ($0.1 - 1)$\,c
within $\leq 5$\,pc, cannot be easily explained with purely hydrodynamical jet models. Most interestingly
is the observed shape of the acceleration curve (dashed line in Figure \ref{cyga} (right)), 
very nicely reproduced by the model of 
Vlahakis \& K\"onigl (\cite{Vlahakis04}, and Fig. 1e therein), which use magnetic acceleration in Poynting-dominated jets.\\

\begin{figure}
\begin{minipage}[t!]{0.49\textwidth}{
\hspace*{4mm} \includegraphics[width=0.60\textwidth,angle=0]{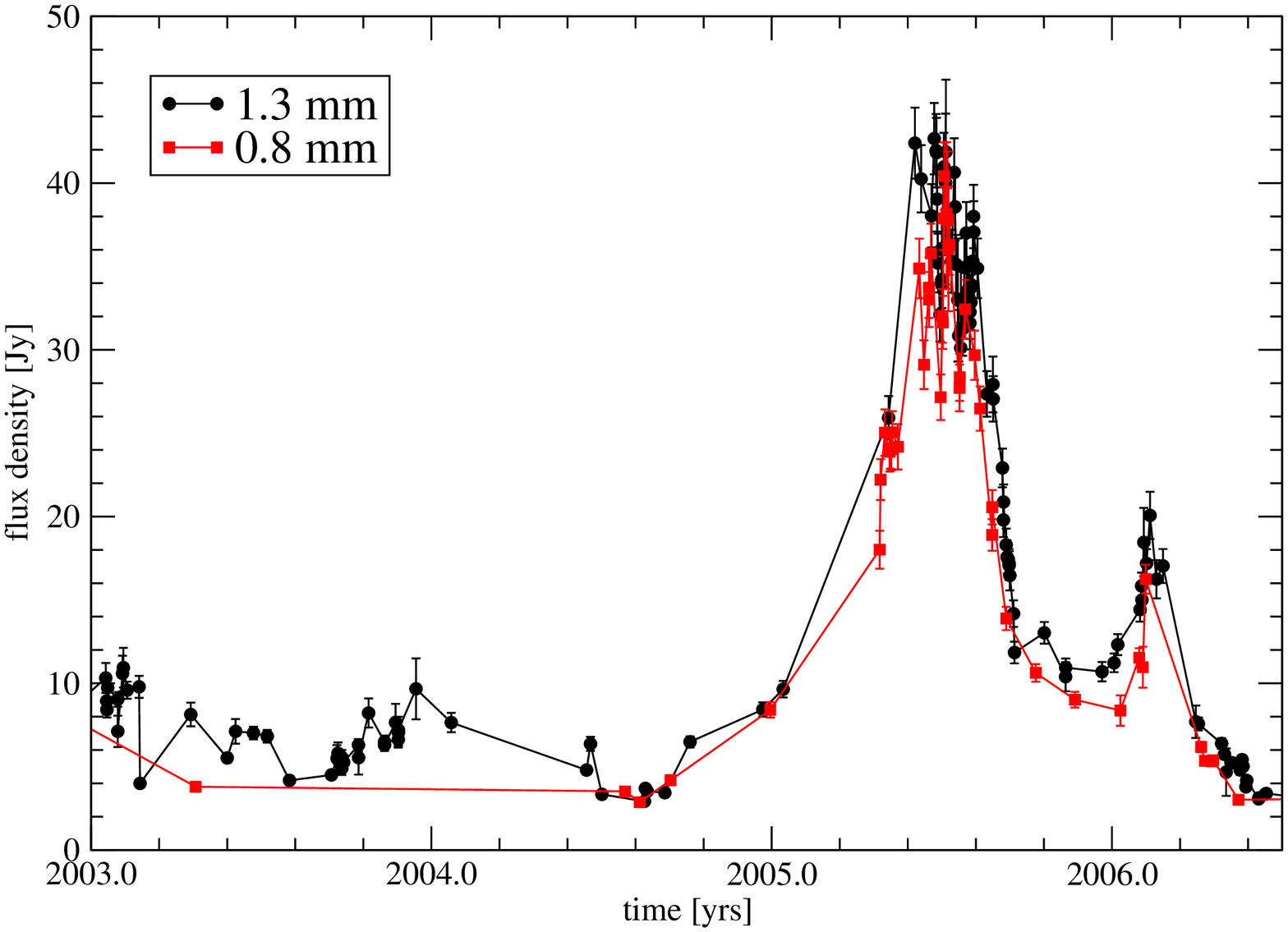}

\hspace*{4mm} \includegraphics[width=0.49\textwidth,angle=-90]{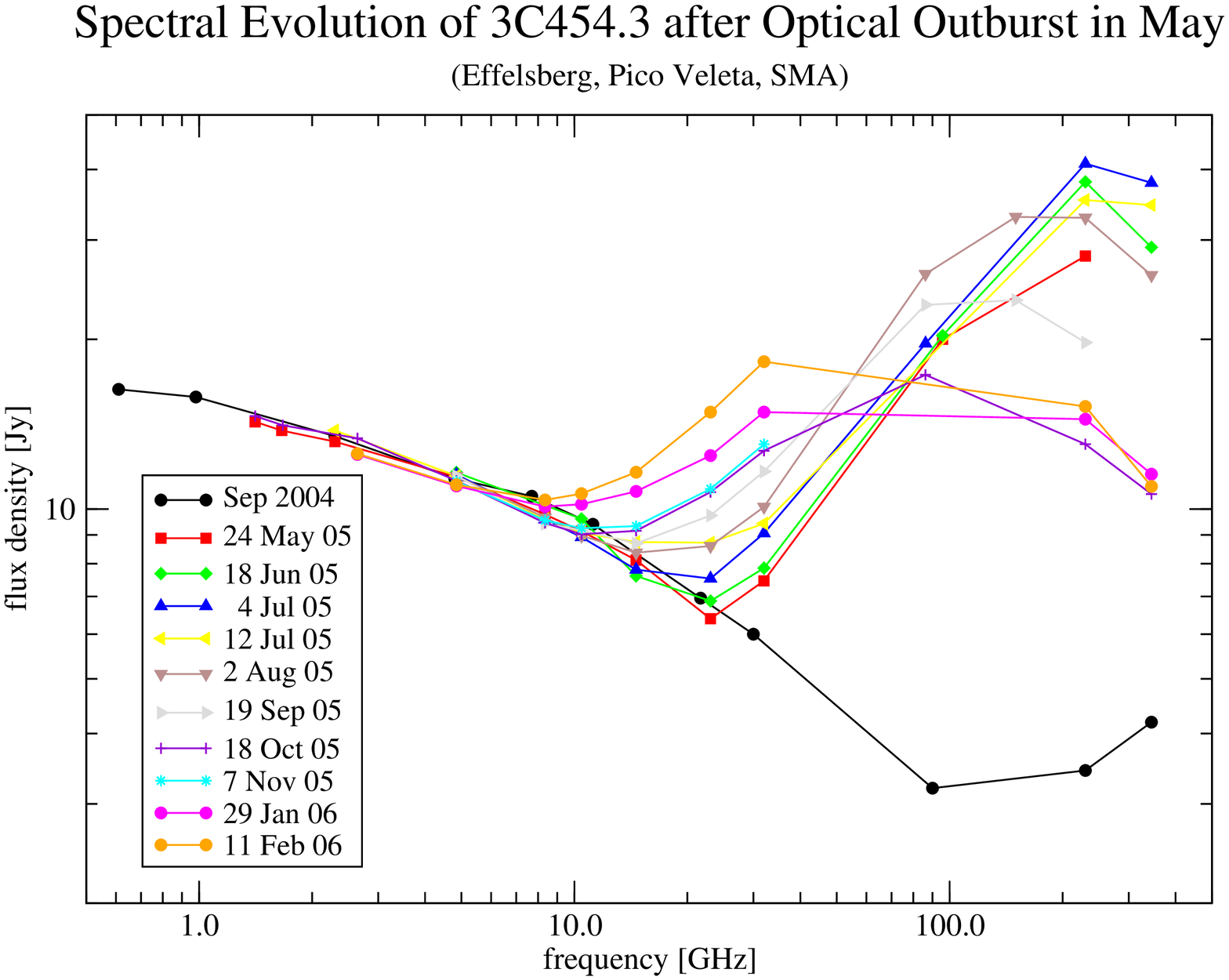}
}
\end{minipage}
\begin{minipage}[t!]{0.49\textwidth}{
\includegraphics[width=0.55\textwidth,angle=0]{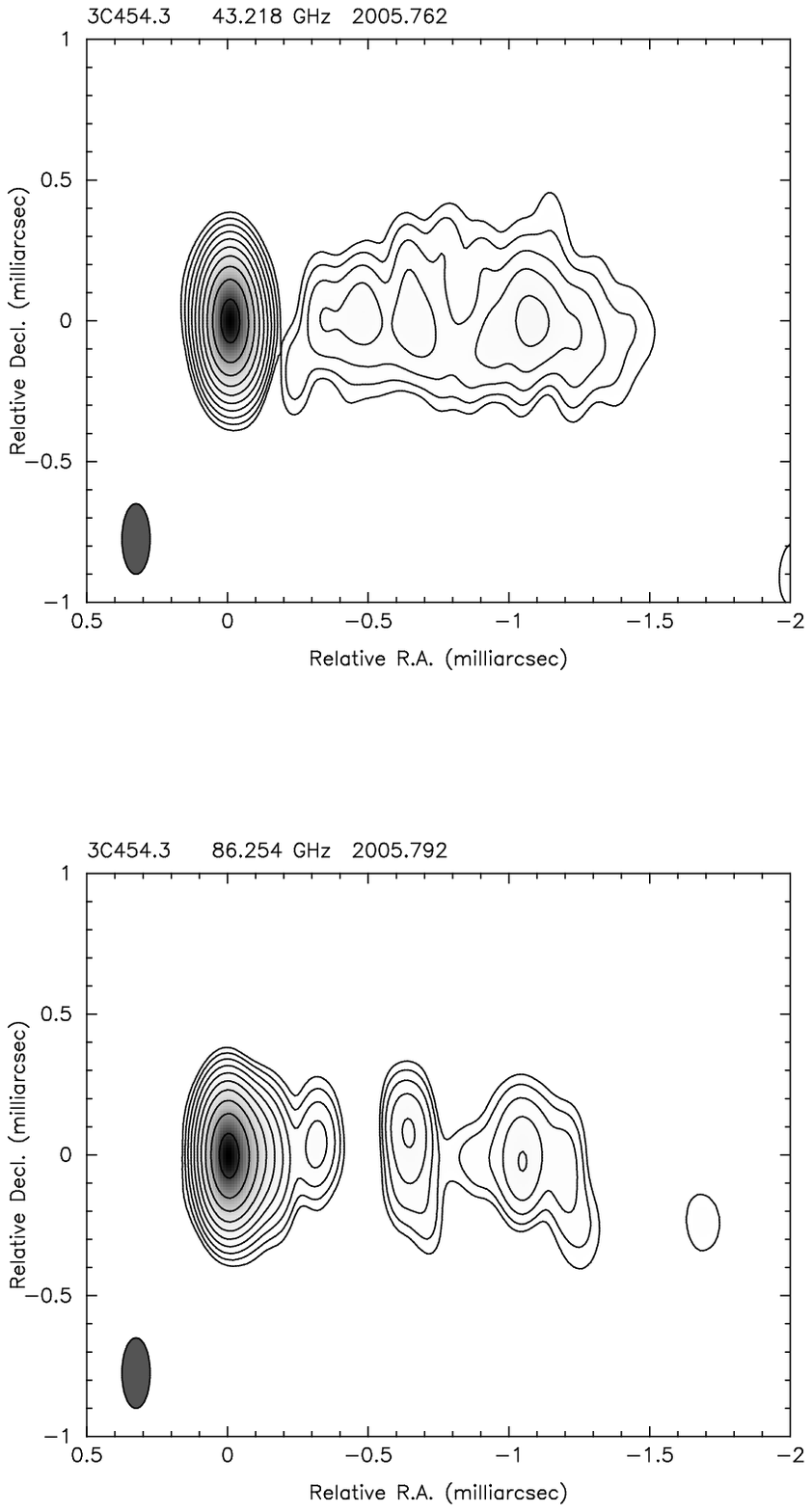}
}
\end{minipage}
\caption{
{\bf Top Left:} The flux density variability of 3C\,454.3 at 1.3\,mm (black) and 0.85\,mm (red). A pronounced
flux density peak is seen around 2005.5. The data are from flux density monitoring performed at the SMA (M. Gurwell, priv. comm.).
{\bf Bottom Left:} Simultaneous radio spectra obtained with a coordinated broad-band spectral monitoring program performed
during 2005--2006.  Data from Effelsberg (1-32\,GHz), Pico Veleta (90-230\,GHz) and from the SMA (230-345\,GHz) are combined.
For each individual observing date, the measurements were done simultaneously within less than $1-3$ days.
Different symbols and colors denote for the different observing epochs.
{\bf Right:} Quasi-simultaneous VLBI images of 3C\,454.3 at 43\,GHz (top, VLBA data: A. Marscher, priv. comm.) and at
86\,GHz (bottom, GMVA) on October 6 and 17, 2005, respectively. Both maps are convolved with
a common elliptical beam of size $0.25 \times 0.1$\,mas. A core elongation observed at the $r=0.1-0.3$\,mas scale
at 86\,GHz is not visible at 43\,GHz, indicative of strong absorption effects in this jet region.
}
\label{flare}
\end{figure}

\n
{\bf 3C\,454.3:}
In AGN, flux density outbursts usually are more pronounced at higher radio frequencies. 
Quite often, they lead to the ejection/creation of new jet components, which subsequently propagate down the jet.
One particular advantage of mm-VLBI is its ability to detect these new jet components in their earliest 
evolutionary phases and near their origin. In early May 2005, the quasar 3C\,454.3 (z=0.859), showed a large flux density outburst,
which was first observed in optical/X-ray bands (\cite{Fuhrmann06},\cite{Giommi06} and references therein) 
and within one month later evolved into a huge millimeter
flare of peak flux of $\sim 40$\,Jy at 3\,mm and 1\,mm wavelength (Fig. \ref{flare}, top left). Already during the onset
of the radio flare, we started to observe the source with a broad frequency coverage and the aim to monitor
the evolution of the radio spectrum. At the 100\,m MPIfR radio telescope
we performed a regular flux density monitoring, rapidly switching between all available receivers (1.4 - 32\,GHz)
on each observing date.
These data are complemented by quasi-simultaneous measurements with the 30\,m IRAM telescope on Pico Veleta 
($90-230$\,GHz, data: H. Ungerechts et al.) and the Sub-millimeter Array (SMA) on Mauna Kea ($230 - 345$\,GHz, 
data: M. Gurwell et al.). In Fig. \ref{flare} (bottom, left) we show some examples of the resulting broad-band 
radio spectra, covering a time range from May 2005 to February 2006. For comparison, also a quiescent pre-flare
spectrum from RATAN-600 is added (Sep. 2004, black line, data: Trushkin et al., \cite{Trushkin05}).
The figure clearly shows a spectral `bump', initially peaking near $\sim 230$\,GHz, and then fading and propagating towards
longer wavelengths. This behavior appears largely consistent with a moving and expanding relativistic shock, 
undergoing synchrotron and adiabatic cooling (e.g. \cite{MarscherGear85}). We note that this
flare appears very pronounced at short millimeter wavelengths, but at longer wavelengths (below $\sim 10$\,GHz) 
so far did not cause very dramatic variations. Therefore it is expected that most of the flaring and related
possible structural variability should occur in the -- at cm-wavelength self-absorbed -- core region.

3C\,454.3 is also monitored with the VLBA and the GMVA by several groups. Unfortunately, no 3\,mm VLBI image
is yet available at a time near the flare maximum (2005.5). In Figure \ref{flare} (right, bottom), we show a
3\,mm GMVA image obtained $\sim 3$ months afterwards (Oct. 17, 2005). The position of a new jet component
seen at $r=0.1$\,mas, is consistent with an assumed ejection between 2005.3 and 2005.5 and an apparent speed
in the range of $5-14$\,c. 

For comparison, a 43\,GHz VLBA image (data: A. Marscher, priv. comm.) observed 11 days earlier is 
shown on top of the 3\,mm map. Both maps show the one sided inner jet, 
extending to $r \simeq 1.5$\,mas length, with at least 3 embedded more compact emission components. 
For the compact, and at 3\,mm slightly elongated core, we find a spectral index of 
$\alpha_{\rm 43/86\,GHz} = -0.4$ ($S \propto \nu^\alpha$). 
While the 86\,GHz map shows two secondary jet components on the $0.8-2.3$\,pc scale near the core
(at $0.1 \leq r \leq 0.3$\,mas), the 43\,GHz image does not show similar emission. This could indicate either
unlikely fast motion within 11 days (of several mas/yr), or the presence of internal absorption. 
By integrating the flux in this region, we estimate a highly inverted spectral index in the range of 
$\alpha_{\rm 43/86\,GHz} = +1.1 ... 2.8$. At larger core separation, however, the spectrum of the jet is steep again.
At present, it is unclear, if this spectral inversion is due to a 
fore-ground absorber partially covering the inner jet. Future polarization and Faraday rotation measurements
could help in answering this question.\\

\begin{figure}
\includegraphics[width=0.95\textwidth,angle=0]{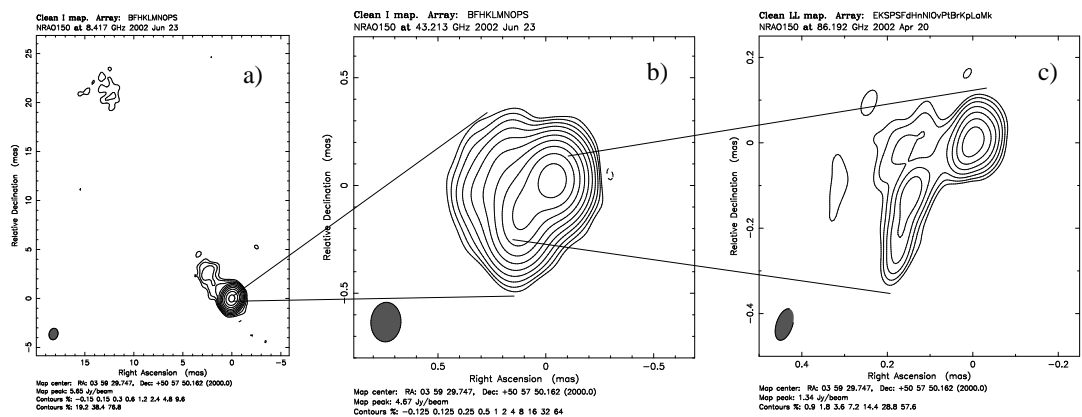}
\caption{The VLBI jet of NRAO\,150. Three maps observed between April and June 2002 at three different frequencies
are shown: 8.4\,GHz (a), 43.2\,GHz (b) and 86.2\,GHz (c). The misalignment and bending of the jet between
milliarcsecond and sub-milliarcsecond scales is obvious. Data: Agudo et al. 2005, and in prep..}
\label{nrao150maps}
\end{figure}

\begin{figure}
\begin{minipage}[t!]{0.49\textwidth}{\hspace*{2mm}
\includegraphics[width=0.9\textwidth,angle=0]{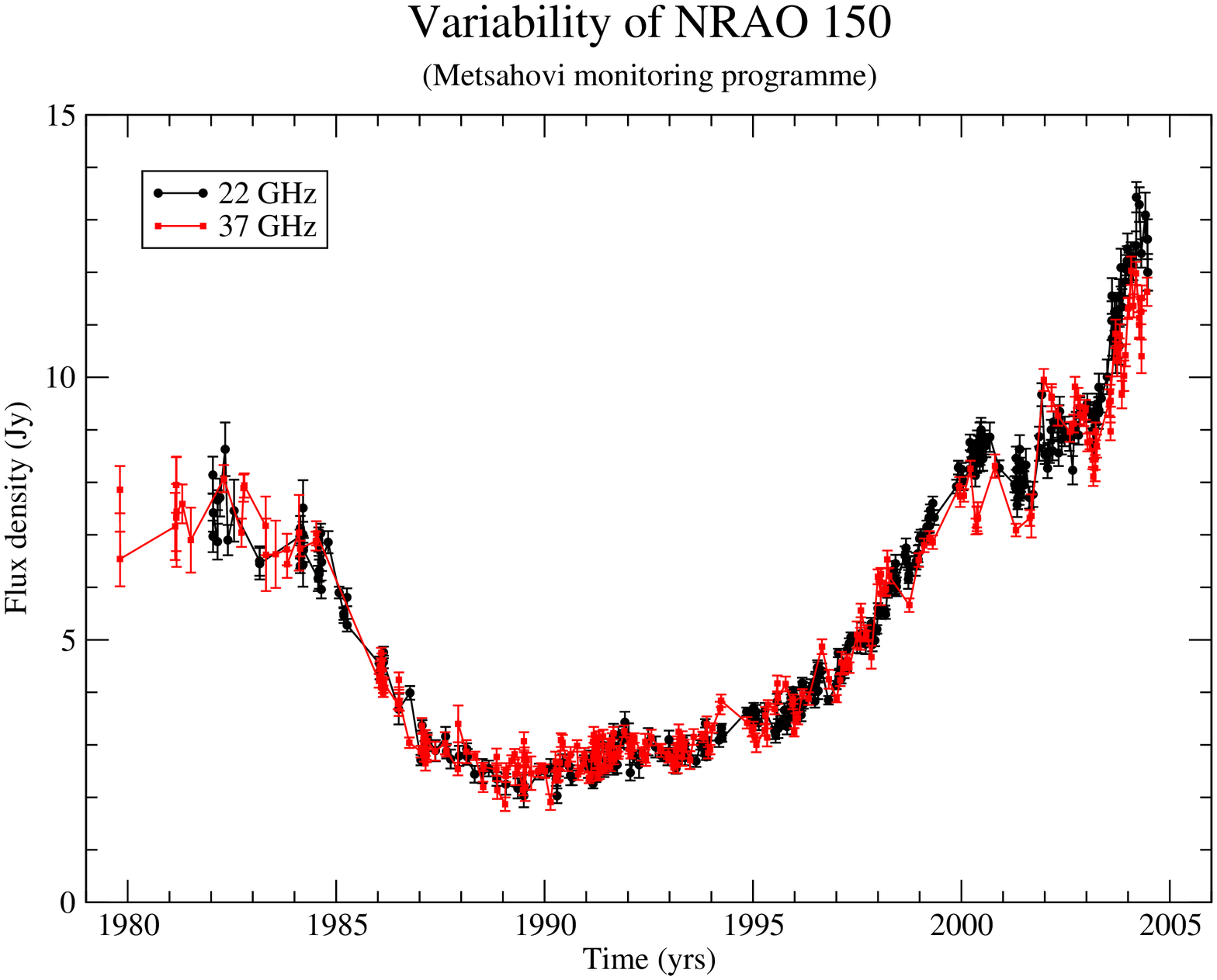}
}
\end{minipage}
\begin{minipage}[t!]{0.49\textwidth}{\hspace*{3mm} \vspace*{-8mm}
\includegraphics[width=0.8\textwidth,angle=0]{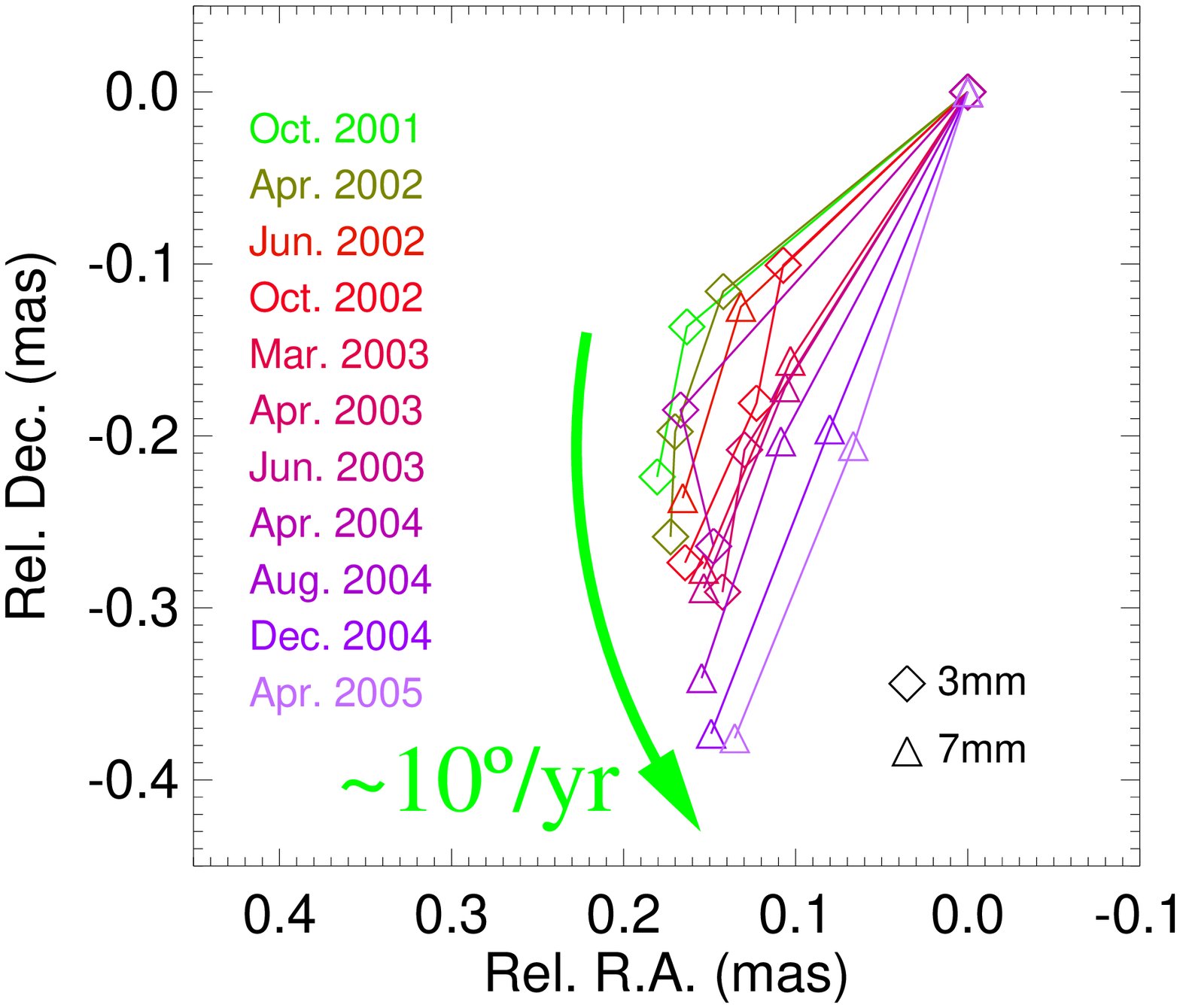}
}
\end{minipage}
\caption{{\bf Left:} Long term flux density variability of NRAO\,150 at 22\,GHz (black) and 37\,GHz (red). Data are
from the Mets\"ahovi monitoring program \cite{Terasranta05} and from M. Tornikoski (priv. comm.).
{\bf Right:} The orientation of the jet axis of NRAO150 as function of time. Diamonds (from 86\,GHz) and triangles (from 43\,GHz)
mark relative positions of Gaussian model fit components with respect to the stationary assumed VLBI core. The lines
connecting adjacent jet components illustrate the orientation of the jet ridge-line and its variation with time.
Different colors denote for various observing epochs (October 2001 to April 2005). Data: Agudo et al., \cite{Agudo05}, \cite{Agudo07}.}
\label{paswing}
\end{figure}

\n
{\bf NRAO\,150:} The last important aspect, which we like to address in this paper, is the possibility
to study with mm-VLBI the nature of jet bending and trace strongly curved jets 
down to their physical origin. In almost all sources so far imaged by mm-VLBI, this jet curvature increases
with decreasing core separation.  An exciting example for a curved and possibly `rotating' jet nozzle is seen in the
strong radio source NRAO\,150 ($z \simeq 1.52$, \cite{Acosta07}). It shows a `sinusoidally' appearing long-term light curve
(Michigan and Mets\"ahovi flux monitoring programs), and a remarkable flux density increase from $\sim 2$\,Jy
to $\sim 12$\,Jy at 37\,GHz during the last 15\,yrs (\cite{Terasranta05}, see also Fig. \ref{paswing}, left). 
Motivated by the possibility of a geometrical origin of this variation and a possible periodicity,
we started to observe the source with global 3\,mm VLBI in the late 1990's (for details see \cite{Agudo05}). 
In Figure \ref{nrao150maps}
we show three VLBI maps obtained at 8, 43, and 86\,GHz in spring 2002. At cm-wavelengths, a one-sided core jet structure
with a jet extending to at least $r = 20$\,mas and oriented along $PA \simeq (30-40)^\circ$ is seen.
At higher angular resolutions, the mm-VLBI images reveal a very different jet orientation of $PA \simeq 140^\circ$,
showing very pronounced jet bending of $\Delta PA \simeq 110^\circ$ within only 0.5\,mas (4.3\,pc) core separation.
Such strong curvature can be explained by geometrical projection effects, if the inner jet is aligned at a small viewing angle. 
43\,GHz VLBA and 86\,GHz GMVA monitoring observations performed during 1999 and 2005 
allow to follow the motion of the inner-most jet components ($\beta_{app} \simeq 2.7 - 3.5$) and measure 
the orientation of the axis of the inner jet. In Figure \ref{paswing} we show a preliminary plot of the 
positions of the inner jet components as function of time and aligned with respect to the stationary assumed VLBI 
core.  Colored lines connect corresponding components at a given epoch 
and outline the orientation of the mean jet axis. A systematic clock-wise rotation of the mean jet 
position angle in the plane of the sky towards
the south, at a rate of $dPA/dt \sim (7-10)^\circ$/yr is obvious. It is tempting to relate this 
rotation to the long term flux density variability, which suggests a timescale 
of $P \gsim 20-25$\,yrs. We note that similar, although 
less pronounced jet position angle variations are seen in the pc-scale jets
of an increasingly large number of sources (i.e. 3C\,84, 3C\,120, 0716+714, OJ\,287,
3C\,273, 3C\,345, BL\,Lac, ...).
This suggests a common physical origin and a fundamental process in AGN radio jets.
We stress that the non-ballistic motion of the jet components excludes simple geometrical precession, as 
is observed in micro-quasars like e.g. SS\,433.  Whether such `wobbling' of the jet foot-point is caused by 
gravitational interaction with another super massive body, by instabilities in the accretion disk or at the jet base,  
or is an inherent property of a BH-accretion disk system (general relativistic effects, e.g. the Lens Thirring rotation), 
remains at this point an open question and deserves future studies.

\begin{figure}[t!]
\begin{minipage}[t!]{1.5\textwidth}{
\includegraphics[width=0.25\textwidth,origin=c]{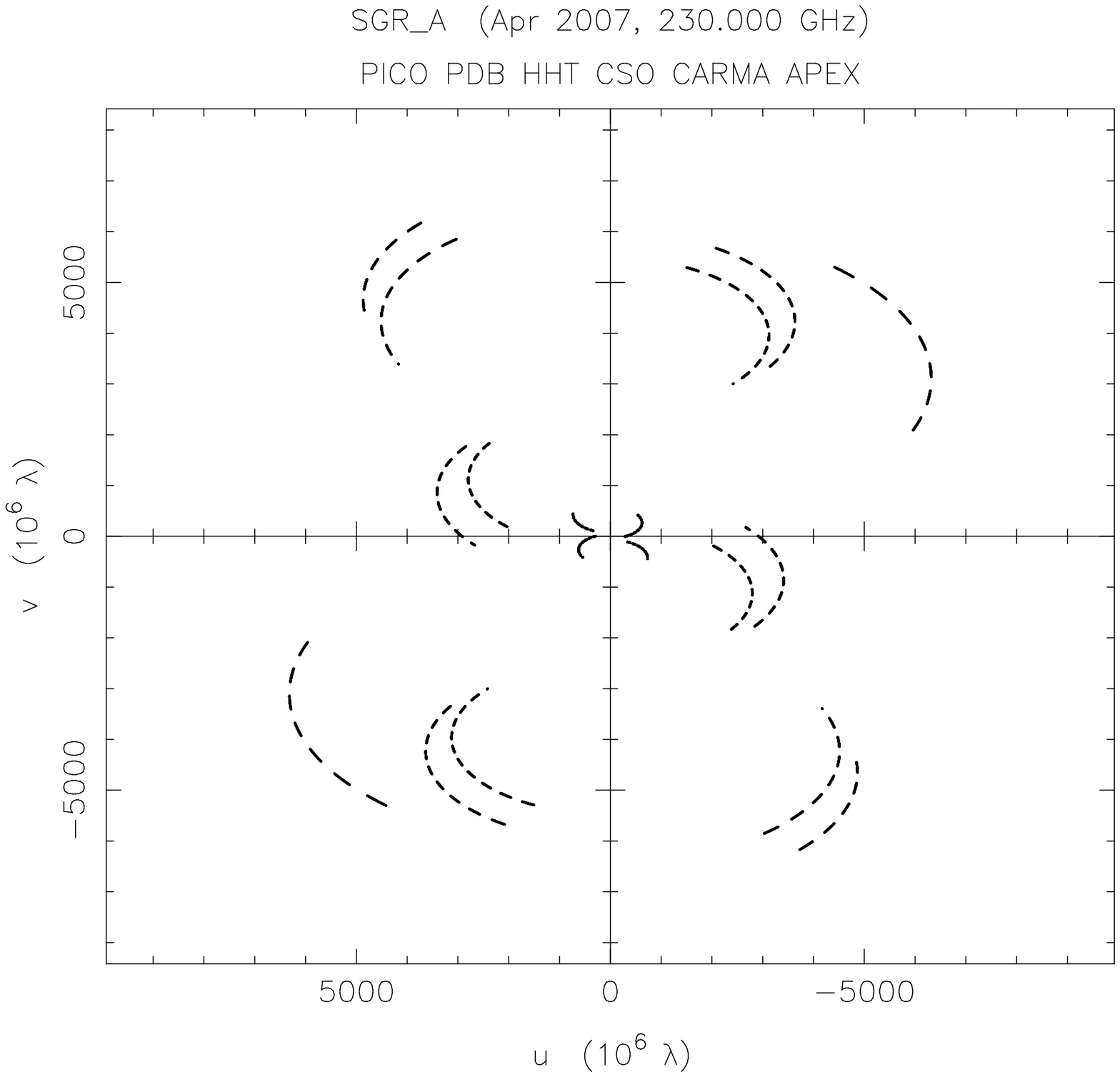}
\hspace*{5mm}\includegraphics[width=0.22\textwidth,angle=-90,origin=br,bb=50 24 268 383,clip=]{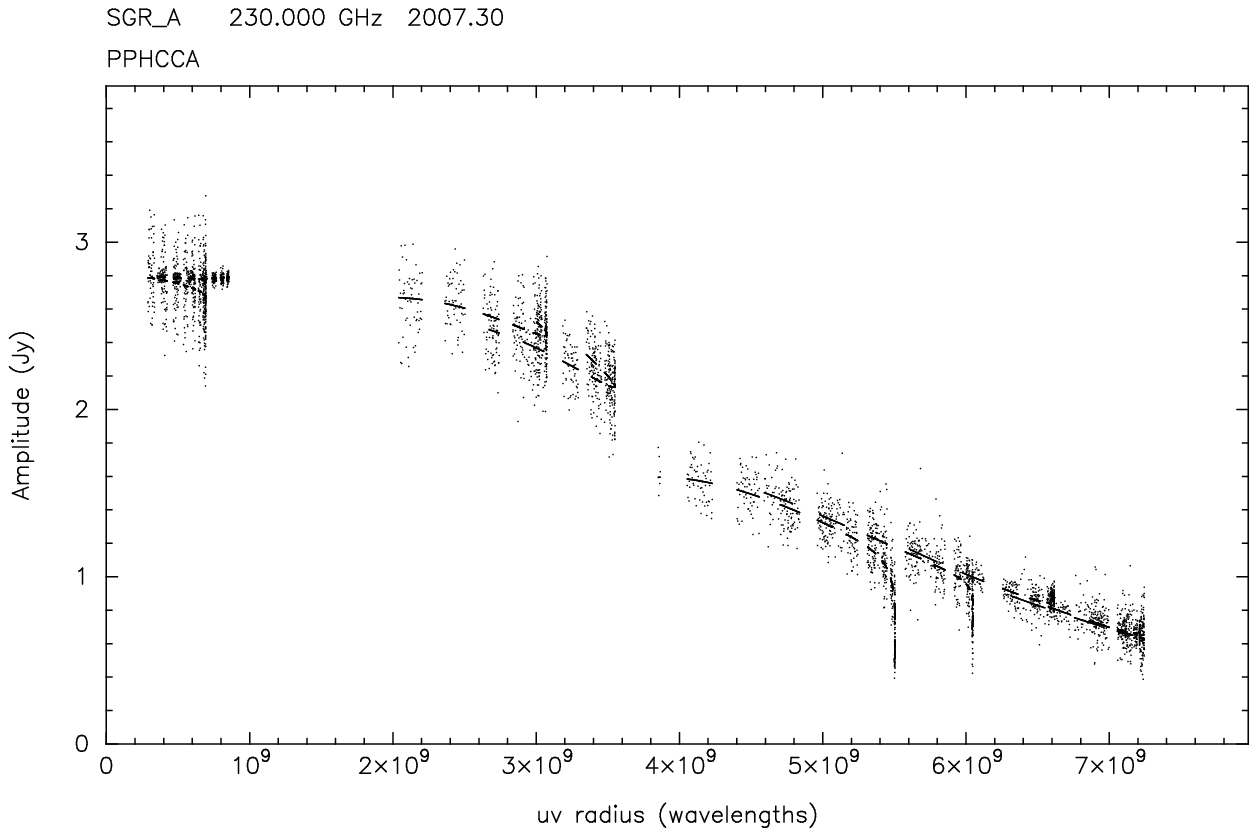}
}
\end{minipage}
\caption{The uv-coverage (left) and visibility amplitude plotted versus projected baseline length (right) of a future
and not unrealistic global VLBI experiment at 1.3\,mm (230\,GHz). For this simulation the participation of the following stations
was assume: Pico Veleta (30\,m, IRAM, Spain), Plateau de Bure (6x15\,m, IRAM, France), Heinrich Hertz Telescope (10\,m, ARO,
Arizona), CSO (10\,m, Caltech, Hawaii), CARMA (single 10.4\,m, California), APEX (12\,m, MPIfR, Chile). For the simulations
we assumed the brightness distribution as shown in the Figure 9 below.
}
\label{1mmsim}
%\end{figure}
~~~~\\
%\begin{figure}
\begin{minipage}[t!]{1.5\textwidth}{
\includegraphics[width=0.25\textwidth,bb=42 150 575 684,clip]{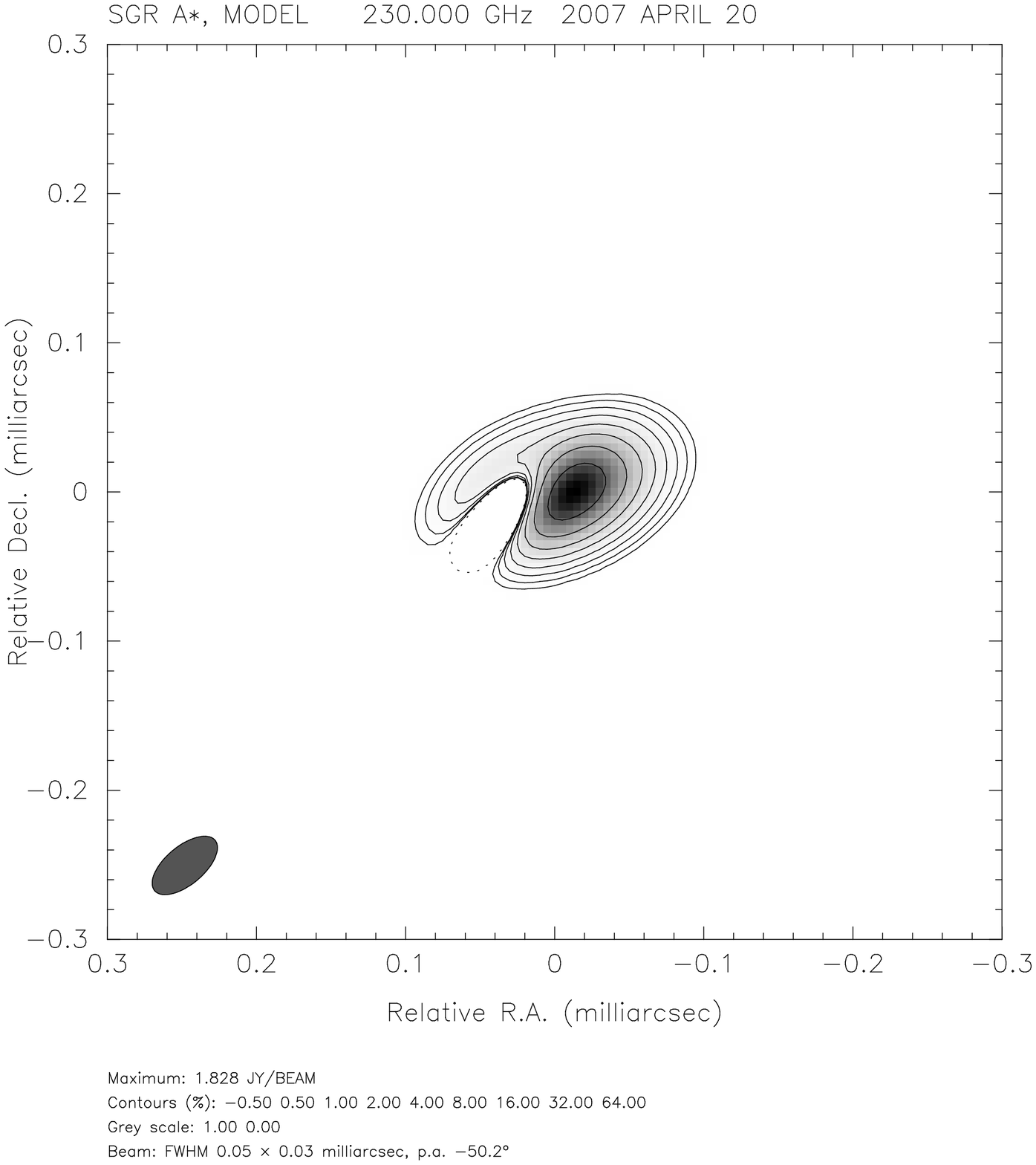}
\hspace*{10mm}\includegraphics[width=0.27\textwidth,angle=-90,origin=br,bb=35 196 470 610,clip]{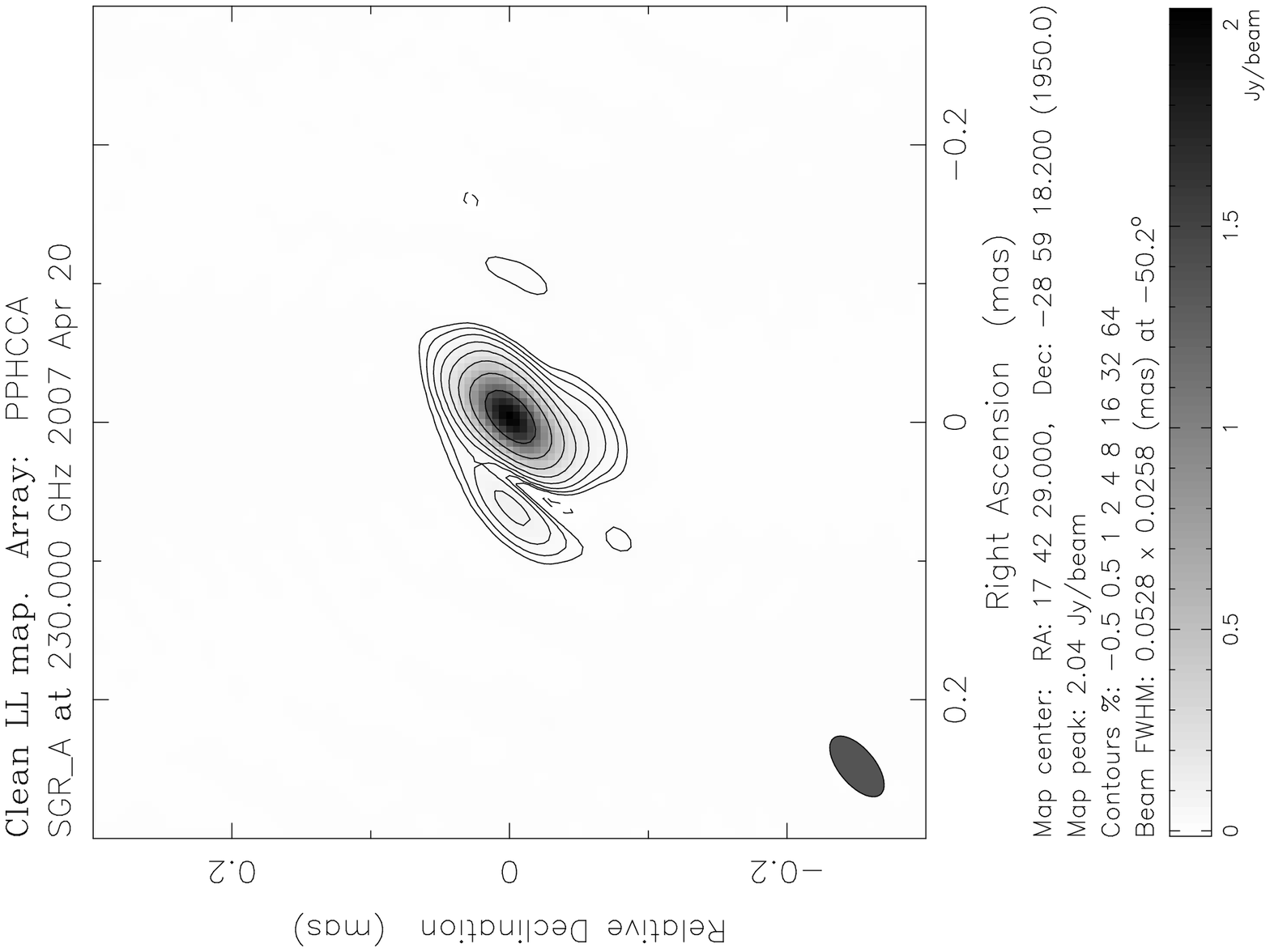}
}
\end{minipage}
\caption{{\bf left:} Hypothetical brightness distribution of an eccentrically displaced sickle-shaped structure as
expected from gravitational light bending around a rotating BH. 
Contour levels are at -0.5, 0.5, 1, 2, 4, 8, 16, 32, and 64\,\% of the peak flux of 1.8\,Jy/beam.
The model is convolved with a restoring beam of size $50 \times 30$\,$\mu$as, oriented at $pa=-50^\circ$. 
{\bf Right:} Reconstructed CLEAN image from the simulated visibility data of Figure 8. Contour levels are the same
as for the input model. The uniformly weighting restoring beam is 
$53 \times 26$\,$\mu$as, oriented at $pa=-50^\circ$.
}
\label{sgramaps}
\end{figure}

\section{Summary and Outlook}

Global millimeter VLBI provides images of dynamic range of a few hundred and with angular resolution
of up to $40-50$\,$\mu$as. Although the sensitivity and uv-coverage is already quite good, there is still
room for some improvement. The performance of the VLBA antennas at 86\,GHz is still limited
by remaining pointing and focusing problems at this high frequency. The lack of a large millimeter antenna
in the USA, limits the transatlantic detection threshold to $\geq 50-100$\,mJy (IRAM-VLBA) 
and the inner-American baseline sensitivity 
to $\geq 200-300$\,mJy. The participation of a large telescope like the GBT in 3\,mm VLBI, would improve
the overall array sensitivity by at least a factor of 2 from $\simeq 2$\,mJy/hr to $\lsim 1$\,mJy/hr. 
A similar argument applies for CARMA, when used as a phased array in 3\,mm-VLBI.
A substantial 
gain in sensitivity comes from a larger observing bandwidth. Whereas European antennas
can now already observe at data rates as high as 1\,Gbit/s (256\,MHz bandwidth for 2 bit sampling), the VLBA
does not yet support this observing mode. 
The application of more sophisticated fringe fitting methods, i.e. by the use of phase 
corrections from water vapor radiometry (see paper by A. Roy et al., this proceeding) and the enhancement of
the AIPS fringe fitting tasks through more sophisticated incoherent averaging 
methods (\cite{Rogers95}) could, further improve the sensitivity of mm-VLBI. 

The mm-VLBI imaging of nearby sources (such as Sgr\,A*  and M\,87)
opens a challenging perspective of being able to study the immediate vicinity of super massive black
holes with a spatial resolution of only a few to a few ten gravitational radii. The low declination of
these particular sources, unfortunately limits the north-south resolution of the existing mm-VLBI networks.
For the imaging of general relativistic effects in nearby black holes,
the addition of mm-VLBI capable radio telescopes in the southern hemisphere (South America: APEX, LMT, ALMA, ...
for observations at $\lambda \leq 3$\,mm), but also in Southern Europe (Noto, Yebes, Sardinia Radio Telescope,
for observations at $\lambda=7 \& 3$\,mm) and in South Africa (no mm-telescope planned)
would be extremely important. To demonstrate this, we show in Figure \ref{1mmsim} and \ref{sgramaps}
a simulation of a future 1.3\,mm VLBI experiment with 6 sub-millimeter telescopes, with one telescope located
in the southern hemisphere (here we used APEX (12\,m), but other antennas like the Japanese ASTE (10\,m) or
the first ALMA antennas may be used, if equipped for VLBI). The simulation shows that a possible intrinsic asymmetry
in the brightness distribution of Sgr\,A* could be detected, if one of the afore mentioned antennas would be
equipped for VLBI (H-maser, VLBI terminal). For this simulation we assumed state of the art receiver temperatures and
a moderate recording rate of 2\,Gbit/s, which reflects the ongoing development of the VLBI data acquisition. Even
higher data rates (observing bandwidths) are anticipated in the next few years (e.g. Mark5 B,
recording at 4 Gbit/s, \cite{Whitney06}). The corresponding sensitivity increase will lead to a
much larger number of observable objects and to mm-VLBI at even shorter wavelengths ($\lambda \leq 1.3$\,mm).
With an angular resolution of better than $10-20$\,$\mu$as one can image compact emission regions of a few gravitational
radii size. Similar small scales can be reached with future space VLBI missions at longer wavelengths
(e.g. VSOP-2, H. Hirabayashi, this conference), and with X-ray interferometers (e.g. MAXIM, \cite{White00}).
The combination of high resolution radio interferometry and interferometric X-ray spectroscopy would indeed
form an extremely powerful tool to study super massive black holes and their environment with unprecedented
accuracy.

\acknowledgments
We thank the staff of the observatories participating in mm-VLBI, and in particular the telescope and
correlator operators and VLBI friends. We thank W. Alef and D. Graham for their help in the various stages
of data reduction.
The VLBI maps presented for 3C\,120 are obtained within a collaboration of some of the authors of this paper with
J.L. G\'omez and A. Marscher. The VLBI results presented for 3C\,454.3 are
obtained partly within a larger collaboration of some of the authors with Boston University Group (A. Marscher et al.), 
the Finish group at Tuorla Observatory (T. Savolainen, K. Wiik, et al.), and IRAM (H. Ungerechts, H. Wiesemeyer, C. Thum et al.).
We especially would like to thank A. Marscher for providing data prior to publication.
For the use of their partly unpublished flux density monitoring data, we thank M. Gurwell (SMA), H. Ungerechts
(IRAM) and M. Tornikoski (MRO). I. Agudo acknowledges financial support from the
European Commission for Science and Research through the ENIGMA network (contract HPRN-CT-2002-00321).

\end{document}